\documentclass[twocolumn,superscriptaddress,showpacs,aps,pra]{revtex4-1}
\usepackage{ulem,bm}
\usepackage{amsmath, upgreek, amssymb, graphicx, paralist, gensymb,epstopdf}
\usepackage[colorlinks, linkcolor=blue, citecolor=blue, urlcolor=blue]{hyperref}
\usepackage{array,epsfig,amsmath,amssymb}
\usepackage{graphicx} 
\usepackage{dsfont}
\usepackage{color}

\def\bra#1{\langle #1|}
\def\ket#1{|#1 \rangle}
\def\bracket#1#2{\langle #1|#2 \rangle}

\begin{document}

\title{Quantum teleportation is a reversal of quantum measurement}

\author{Seung-Woo Lee}
\email{swleego@gmail.com}
\affiliation{Center for Quantum Information, Korea Institute of Science and Technology, Seoul, 02792, Korea}

\author{Dong-Gil Im}
\affiliation{Department of Physics, Pohang University of Science and Technology, Pohang, 37673, Korea}

\author{Yoon-Ho Kim}
\affiliation{Department of Physics, Pohang University of Science and Technology, Pohang, 37673, Korea}

\author{Hyunchul Nha}
\affiliation{Department of Physics, Texas A\&M University at Qatar, Education City, P.O.Box 23874, Doha, Qatar}

\author{M. S. Kim}
\affiliation{QOLS, Blackett Laboratory, Imperial College London, London, SW7 2AZ, United Kingdom}
\affiliation{Korea Institute for Advanced Study, Seoul, 02455, Korea}

\date{\today\\}

\begin{abstract}

We introduce a generalized concept of quantum teleportation in the framework of quantum measurement and reversing operation. 
Our framework makes it possible to find an optimal protocol for quantum teleportation enabling a faithful transfer of unknown quantum states with maximum success probability up to the fundamental limit of the no-cloning theorem. 
Moreover, an optimized protocol in this generalized approach allows us to overcome noise in quantum channel beyond the reach of existing teleportation protocols without requiring extra qubit resources. 
Our proposed framework is applicable to multipartite quantum communications and primitive functionalities in scalable quantum architectures.
\end{abstract}
\maketitle

\section{Introduction}

Quantum teleportation is at the core of quantum information technologies \cite{duan01,Gottesman99, Knill01,Pirandola15}. Ever since the seminal work by Bennett {\em et al.}~\cite{Bennett93}, teleportation has been demonstrated in various physical platforms \cite{Bouwmeester97,Boschi98,furu,Kim01,Barrett04,Olmschenk09,Steffen13,Pfaff14,wang15,Graham15} and many different protocols have been proposed to date \cite{Nielsen96,Horodecki99,Banaszek00,Karlsson98,Hillery99,Zhao04,Yonezawa04,Barasi19,JLee00,Zhang06,Mor99,WSon00,Roa03,Li00,Agrawal02}. Recently, developments have been directed to its potential applications for long-distance quantum communications \cite{Riedmatten04,Sum16,Valivarthi16,Ren17, Valivarthi20}, inter-chip communications \cite{Llewellyn20} and other versatile functionalities toward scalable quantum architectures \cite{Chou18,Wan19,Lee20}. Teleportation makes it possible to transfer an unknown quantum state $\rho$ from a sender (Alice) to a receiver (Bob) using an entangled quantum channel. In this protocol, Alice performs a joint measurement on an input state $\rho$ and her part of the quantum channel, which results in a reduced state $\rho^i$ at Bob's station under the measurement outcome $i$ at Alice station. To complete the teleportation, Bob applies a conditional operation according to the outcome $i$ aiming at $\rho^i \rightarrow \rho$. 

A crucial question may arise here: For a given a quantum channel and Alice's choice of joint measurement, what is the optimal operation for Bob to recover the input state $\rho$? An ideal teleportation via a maximally entangled quantum channel transfers the information of input state deterministically and faithfully, in which Bob's optimal operation is unitary \cite{Bennett93}. In reality, however, teleportation is frequently affected by imperfections and noise that cause a leak of information and the fidelity of the teleported state is inevitably degraded.
In this respect, applying a unitary operation at the receiver's station may not always be optimal in practice. However, existing teleportation protocols to date conventionally limit Bob's operation to unitary \cite{Nielsen96,Horodecki99,Banaszek00,Karlsson98,Hillery99,Zhao04,Yonezawa04,Barasi19,JLee00,Zhang06,Mor99,WSon00,Roa03,Li00,Agrawal02}, which would also significantly limit the fault-tolerance of quantum teleportation.

In this work, we develop a general framework to optimize the teleportation fidelity beyond the reach of existing teleportation protocols. The general framework is established from the perspective of quantum measurement and reversing operation. This makes it possible to optimize not only the sender's joint measurement but also the receiver's operation. The importance of our work is twofold. 
(i) Our work extends and generalizes the concept of quantum teleportation beyond existing protocols. We show that the optimized protocol in this framework enables a faithful transfer of quantum states with unit fidelity and the success probability up to the limit given by the no-cloning theorem \cite{Cheong12,Lim14}. This can be further extended to multipartite quantum communications.
(ii) Our framework allows us to transfer quantum states over noisy quantum channels, overcoming the limit of existing teleportation protocols. We show that quantum teleportation is possible even through such a decohered entangled channel that is useless in conventional protocols. This may be achieved in a probabilistic way. However, the current technology can generate an entangled channel at a very high repetition rate \cite{Ma20}. As increasing the fidelity is critical for many applications, e.g.~fault-tolerant distributed quantum protocols \cite{Wehner18,Severin21}, a probabilistic scheme may serve well particularly if it works in a heralded fashion. Moreover, in contrast to other methods to cope with noise in quantum channel such as entanglement distillation \cite{bennett96,Bennett97,pan03,yama03} and error correction encoding \cite{Munro12,Muralidharan12,Azuma15,Lee19}, our protocol requires no additional qubits but only to modify the sender's or receiver's operation. Our work constitutes a feasible, resource-efficient, way to mitigate errors while transmitting quantum information.

\section{General framework}

Let us start by introducing a generalized concept of teleportation. We define a quantum teleportation as {\em transmitting an unknown quantum state by disturbing it with measurement at one location and reconstructing it with a certain operation at another location, where two locations are connected via quantum entanglement}.
This not only encompasses all existing teleportation protocols \cite{Bennett93,Nielsen96,Horodecki99,Banaszek00,Karlsson98,Hillery99,Zhao04,Yonezawa04,Barasi19,JLee00,Zhang06,Mor99,WSon00,Roa03,Li00,Agrawal02}  but generalizes the scope of teleportation as we describe below.

Consider a general scenario of one-to-one teleportation. Alice performs a joint measurement $\{\Pi^i\}$ (a positive-operator valued measure, POVM) on the input particle $a$ and her part $\bar{a}$ of the channel state $\Psi_{\bar{a}b}$. The (unnormalized) reduced state can then be written as 
\begin{equation}
\label{eq:stateb}
\tilde{\rho}^i_b=\Pi_{a\bar{a}}^i(\rho_a\otimes\Psi_{\bar{a}b})=M^i_{a\rightarrow b}(\rho).
\end{equation}
Here we define a quantum measurement $\{M_{a\rightarrow b}^i\}$, which is nonlocal in the sense that its input($a$) and output($b$) are spatially separated. 
More specifically, if we assume a rank-one projection $\Pi^i_{a\bar{a}}=\ket{v_i}\bra{v_i}$ and a quantum channel $\Psi_{\bar{a}b}=\ket{\Psi}\bra{\Psi}$ (see Appendix~\ref{asec:eqm} for general scenarios), the reduced state can be written as
\begin{equation}
\label{eq:effm}
\tilde{\rho}^i_b={}_{a\bar{a}}\bra{v_{i}}\cdot\rho_a \otimes \ket{\Psi}_{\bar{a}b}\bra{\Psi}\cdot\ket{v_{i}}_{a\bar{a}}
=\hat{M}_i\rho\hat{M}_i^{\dag},
\end{equation}
with measurement operators $\hat{M}_i\equiv{}_{a\bar{a}}\bracket{v_{i}}{\Psi}_{\bar{a}b}$ satisfying the completeness condition $\sum_i \hat{M}_i^{\dag}\hat{M}_i=\openone$. Therefore, once the outcome $i$ is shared with Bob, the reduced state is equivalent to a post-measurement state obtained when a quantum measurement ${\bf M}=\{ \hat{M}_{i}\}$ is performed on $\rho$ at Bob's location.

\begin{figure}
\centering
\includegraphics[width=0.99\linewidth]{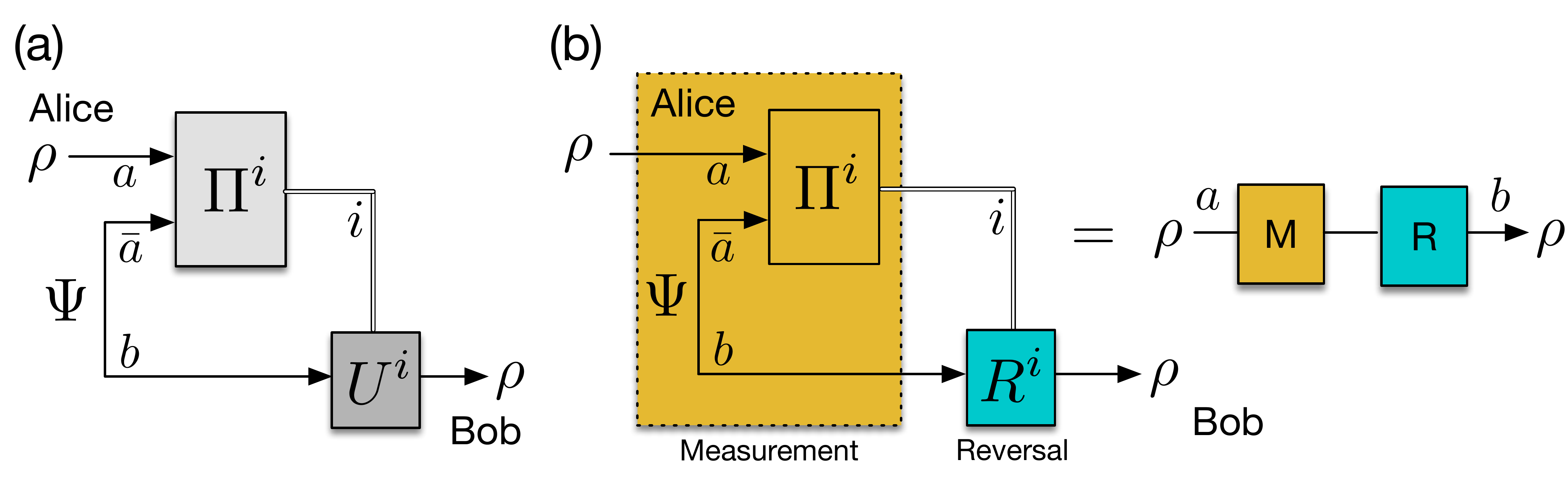}
\caption{(a) Conventional quantum teleportation. (b) Quantum teleportation in the framework of measurement and reversal (MR framework).}
\label{fig:GQT}
\end{figure}

A remaining step to complete the teleportation is applying an appropriate operation to recover $\rho$. Note that this operation has been limited to a conditional unitary operation in previous protocols \cite{Bennett93,Nielsen96,Horodecki99,Banaszek00,Karlsson98,Hillery99,Zhao04,Yonezawa04,Barasi19,JLee00,Zhang06,Mor99,WSon00,Roa03,Li00,Agrawal02}  as illustrated in Fig.~\ref{fig:GQT} (a). In contrast, we envision an operation $\bf R$ to reverse ${\bf M}$ such that
\begin{equation}
\label{eq:GF}
({\bf R}\circ{\bf M}) (\rho) \propto \rho
\end{equation}
 irrespective of $\rho$, where $\propto$ implicates that the process can be probabilistic.
For each outcome $i$ of ${\bf M}$, Bob can choose an appropriate operation ${\bf R}=\{\hat{R}^i_j\}$ to reverse the effect of measurement ${\bf M}$ aiming at a final output state
\begin{eqnarray}
\tilde{\rho}^i_j = \hat{R}^i_{j}\hat{M}^i \rho\hat{M}^{i \dag}\hat{R}_{j}^{i \dag}\rightarrow \rho.
\end{eqnarray}
Note that this framework allows to modify both Alice's joint measurement and Bob's reversing operation, and if the reversing operation is limited to unitary, i.e., $\hat{R}^i=\hat{U}^i$, it reduces to a conventional teleportation.
We put forward the idea that a teleportation protocol is conceptually equivalent to a quantum measurement followed by a reversing operation applied to an unknown quantum state, referred to as {\em Measurement-Reversal (MR) framework} hereafter. 

This approach is generally valid even under the effect of noise. Assume an arbitrary noise affecting the quantum channel (on particle $b$ for simplicity) represented by a set of operators ${\cal E}=\{\hat{E}_k\}$. The reduced state (\ref{eq:effm}) then reads
\begin{equation}
\label{eq:efmn}
\tilde{\rho}^i_b={}_{a\bar{a}}\bra{v_{i}} \rho_a \otimes \sum_k \hat{E}_{k,b} \ket{\Psi}_{\bar{a}b}\bra{\Psi} \hat{E}^{\dag}_{k,b}  \ket{v_{i}}_{a\bar{a}}
=\sum _k \hat{M}_{i,k}\rho\hat{M}^{\dag}_{i,k},
\end{equation}
where $\hat{M}_{i,k}\equiv{}_{a\bar{a}}\bra{v_{i}}\hat{E}_{k,b}\ket{\Psi}_{\bar{a}b}$ satisfying $\sum_{i,k} \hat{M}_{i,k}^{\dag}\hat{M}_{i,k}=\bra{\Psi}\sum_i \ket{v_{i}}\bra{v_{i}}\otimes \sum_k \hat{E}^{\dag}_k \hat{E}_k \ket{\Psi}=\openone$ is a coarse-grained quantum measurement that yields the outcome $i$ while the outcome $k$ is hidden. 
Thus, a teleportation over a noisy quantum channel can be addressed in terms of a coarse-grained quantum measurement and reversing operation.

MR framework is also readily extended to multipartite teleportation \cite{Karlsson98,Hillery99,Zhao04,Yonezawa04,Barasi19}. Consider an arbitrary number of participants; $n$ number of senders (who share a quantum state $\rho$ to teleport), $l$ number of intermediators (who relay the information), and $m$ number of receivers, all of whom are connected by a quantum channel $\ket{\Psi}$. Senders perform a joint measurement on their parts of $\rho$ and $\ket{\Psi}$ in the basis $\ket{V_{\vec{v}}}$, where $\vec{v}\equiv(v_1,\ldots,v_n)$ denotes the outcomes of all senders. On the other hand, intermediators perform a local measurement on their parts of $\ket{\Psi}$ in the basis $\ket{W_{\vec{w}}}$, where $\vec{w}\equiv(w_1,\ldots,w_l)$ denotes the all local outcomes. The operator for this collective measurements is given by
\begin{equation}
\label{eq:multiM}
\hat{M}_{\vec{v},\vec{w}}\equiv{} \bra{W_{\vec{w}}}\bra{V_{\vec{v}}}\cdot \ket{\Psi},~~~\sum_{\vec{v},\vec{w}}\hat{M}^{\dag}_{\vec{v},\vec{w}}\hat{M}_{\vec{v},\vec{w}}=\openone.
\end{equation}
Given $\vec{v}$ and $\vec{w}$, receivers can apply appropriate reversing operations $\hat{R}^{{\vec{v},\vec{w}}}$ such that $\hat{R}^{{\vec{v},\vec{w}}}_{\vec{u}}\hat{M}_{\vec{v},\vec{w}}\rho\hat{M}^{\dag}_{\vec{v},\vec{w}}\hat{R}_{\vec{u}}^{{\vec{v},\vec{w}} \dag} \propto \rho$, where $\vec{u}=(u_1,\ldots,u_m)$ denotes the outcomes of the reversing operations and $\sum_{\vec{u}}\hat{R}_{\vec{u}}^{{\vec{v},\vec{w}} \dag}\hat{R}_{\vec{u}}^{{\vec{v},\vec{w}}}=\openone$. 

\section{Performance measures}
The performance of teleportation can be assessed in terms of the teleportation fidelity and the success probability of the protocol.
Consider a protocol described by ${\bf M}=\{\hat{M}_{i,k}\}$ and ${\bf R}=\{\hat{R}^i_j\}$ to teleport an arbitrary (pure) input state $\rho=\ket{\psi}\bra{\psi}$. After applying ${\bf M}$, the post-measurement state for outcome $i$ is 
$\rho^i=\sum _k \hat{M}_{i,k} \ket{\psi}\bra{\psi}\hat{M}^{\dag}_{i,k}/p^i(\psi)$ with its probability $p^i(\psi)=\sum_k \bra{\psi}\hat{M}^{\dag}_{i,k}\hat{M}_{i,k} \ket{\psi}$. Once ${\bf R}$ succeeds, which is designated as $j=1$ without loss of generality, the teleported state is $\rho^i_j=\sum _k \hat{R}^i_{j}\hat{M}_{i,k} \ket{\psi}\bra{\psi}\hat{M}_{i,k}^{\dag}\hat{R}_{j}^{i \dag}/p^i_j(\psi)$
with $p^i_j(\psi)=\sum_k \bra{\psi}\hat{M}_{i,k}^{\dag}\hat{R}_{j}^{i \dag}\hat{R}^i_{j}\hat{M}_{i,k} \ket{\psi}$. The {\em teleportation fidelity} can then be evaluated as the average fidelity between the input and output states from the success event of the teleportation, i.e.,
\begin{equation}
{\cal F}_{\rm tele}=\int d \psi \sum_{i} p^i(\psi) \bra{\psi} \rho^i_{j=1} \ket{\psi}.
\end{equation}
The average {\em success probability} is given by 
\begin{equation}
\label{eq:Ptele}
{\cal P}_{\rm tele} = \int d \psi \sum_{i} p^i_{j=1}(\psi).
\end{equation}

Since any leakage of information to senders or intermediators would degrade the teleportation performance \cite{Banaszek00}, the maximum amount of extractable information on $\ket{\psi}$ by ${\bf M}$ is also an important measure to consider. 
Suppose that Alice attempts to extract information on $\ket{\psi}$ and estimates it as $\ket{\widetilde{\psi}_i}$ upon the outcome $i$ using her knowledge over ${\bf M}$. Then, the {\em information gain} can be evaluated by the overlap between $\ket{\psi}$ and $\ket{\tilde{\psi}_i}$ on average, 
\begin{equation}
\label{eq:gain}
{\cal G}_{\rm Alice}= \int d\psi \sum_{i}\sum_k p^{i,k}(\psi) |\bracket{\tilde{\psi}_i}{\psi}|^2,
\end{equation}
with $p^{i,k}(\psi)= \bra{\psi}\hat{M}_{i,k}^{\dag}\hat{M}_{i,k} \ket{\psi}$  \cite{Bana}. In multipartite teleportations, the information gain may be evaluated as the total amount of information extracted by all senders and intermediators together.

\section{Fundamental limits}
Now, every trial of teleportation can be seen as a process that has a single input $\ket{\psi}$ and two outputs, i.e., $\ket{\tilde{\psi}_i}$ to Alice and $\rho^i_j$ to Bob. As the estimated state $\ket{\tilde{\psi}_i}$ can be imprinted onto unlimited number of copies, this process in turn can be understood as an asymmetric cloning machine ($1\rightarrow N+1$ where $N=\infty$) \cite{Bae06}. Thus, the performance of teleportation is fundamentally restricted by the no-cloning theorem (see Appendix~\ref{asec:FL}). 

To find the upper limit of performance, consider an arbitrary measurement ${\bf M} =\{\hat{M}_{i}\}$ without a noise channel. Each measurement operator can be represented by a singular value decomposition as $\hat{M}_i=\hat{V}_i\hat{D}_i\hat{U}_i$ with unitary operators $\hat{U}_i$ and $\hat{V}_i$, and a diagonal matrix $\hat{D}_i=\sum_{n=0}^{d-1}\lambda^i_n\ket{n} \bra{n}$ with singular values $\lambda^i_n$ in $d$-dimensional Hilbert space. The information gain (\ref{eq:gain}) can then be further evaluated as ${\cal G}^{\rm max}_{\rm Alice}=(d+\sum_{i} (\lambda_{\rm max}^i)^2)/d(d+1)$ in terms of the largest singular values $\lambda^i_{\rm max}$ \cite{Bana} (see Appendix~\ref{asec:pm}), which is scaled in the range $1/d \leq {\cal G}^{\rm max}_{\rm Alice} \leq 2/(d+1)$. We note that its upper bound determines the {\em classical limit of teleportation}, i.e., ${\cal F}_{\rm cl}= 2/(d+1)$. This is because the maximum information gain by Alice is the maximum amount of information that can be used in the classical measure-prepare strategy. On the other hand, the success probability (\ref{eq:Ptele}) can be evaluated as ${\cal P}^{\rm max}_{\rm tele}=\sum_i(\lambda^i_{\rm min})^2$ in terms of the smallest singular value $\lambda^i_{\rm min}$ of each $\hat{M}_{i}$ (see Appendix~\ref{asec:pm}) \cite{Cheong12}. 

Finally, following \cite{Cheong12}, we can find the trade-off relation between the maximum teleportation probability and the information gain by Alice as
\begin{equation}
\label{eq:balance} 
d(d+1) {\cal G}^{\rm max}_{\rm Alice}+(d-1){\cal P}^{\rm max}_{\rm tele} \leq 2d,
\end{equation}
which determines the fundamental limit of teleportation performance. It implies that the more information is extracted during the teleportation, the lower the success probability of the faithful teleportation. Note that the limit (\ref{eq:balance}) generally holds for any teleportation protocols.

\section{Optimal faithful teleportation}

We find that any teleportation protocol for an entangled quantum channel without noise can be optimized to yield the unit teleportation fidelity ${\cal F}_{\rm tele}=1$ with the maximum success probability ${\cal P}^{\rm max}_{\rm tele}$.  
For each measurement operator $\hat{M}_i=\hat{V}_i\hat{D}_i\hat{U}_i$, we can define a reversing operator $\hat{R}^i=\lambda^i_{\rm min}\hat{U}^{\dag}_i\hat{D}^{-1}_i\hat{V}^{\dag}_i$ such that
\begin{equation}
\label{eq:oR}
\hat{R}^i \hat{M}_i |\psi \rangle = \lambda^i_{\rm min} |\psi\rangle,~\forall \ket{\psi}.
\end{equation}
It guarantees that an unknown state $\ket{\psi}$ can be faithfully transferred with success probability ${\cal P}^{\rm max}_{\rm tele}=\sum_i(\lambda^i_{\rm min})^2$.
For instance, consider a teleportation via an entangled channel $\ket{\Psi}=\cos{(\theta/2)}\ket{00}+\sin{(\theta/2)}\ket{11}$ with $0 \leq \theta \leq \pi/2$. 
If we fix the joint measurement on the Bell basis for simplicity (see Appendix~\ref{asec:Comp} for a general scenario), the measurement operator in MR framework corresponds to $\hat{M}_i=(\cos{(\theta/2)}\ket{0}\bra{0}+\sin{(\theta/2)}\ket{1}\bra{1})\hat{U}_i/\sqrt{2}$, where $\hat{U}_i \in \{\hat{I},\hat{\sigma}_z,\hat{\sigma}_x,\hat{\sigma}_{x}\hat{\sigma}_z \}$. Its optimal reversing operator is then given as $\hat{R}^i=\hat{U}_i(\tan{(\theta/2)}\ket{0}\bra{0}+\ket{1}\bra{1})$. Hence, ${\cal F}_{\rm tele}=1$ can be achieved with probability ${\cal P}^{\rm max}_{\rm tele}=2\sin^2(\theta/2)$ that monotonically increases with the degree of entanglement in quantum channel $E=\sin(\theta/2)$. The information gain (\ref{eq:gain}) is obtained as ${\cal G}^{\rm max}_{\rm Alice}=(1+\cos^2(\theta/2))/3$, so ${\cal P}^{\rm max}_{\rm tele}$ and ${\cal G}^{\rm max}_{\rm Alice}$ saturate the upper bound in Eq.~(\ref{eq:balance}). 
This generally holds for arbitrary multipartite teleportation scenarios as we can define a reversing operator to fulfill (\ref{eq:oR}) for the measurement operator ({\ref{eq:multiM}).
As a result, it is clear in MR framework that {\it a faithful teleportation, i.e., ${\cal F}_{\rm tele}=1$, is always possible over a pure entangled channel between arbitrary number of participants, achieving the maximum success probability ${\cal P}^{\rm max}_{\rm tele}$}.

We note that a faithful teleportation optimized in MR framework differs from previously proposed protocols in the context of conclusive \cite{Mor99,WSon00,Roa03} or probabilistic teleportation \cite{Li00,Agrawal02}. Such protocols can be optimized by modifying Alice's joint measurement only, so their implementation as well as extension to multipartite teleportations is nontrivial. By contrast, in MR framework, it is rather straightforward to identify an optimal protocol, even under a noisy environment shown later. A teleportation protocol can be optimized not only by Alice's joint measurement but also through Bob's reversing operation that can be implemented by a single qubit (qudit) operation. Therefore, an optimal teleportation in MR framework is readily implementable and generally applicable to arbitrary multipartite teleportation scenarios. The detailed comparison is given in Appendix~\ref{asec:Comp}.

\section{Overcoming noise in quantum channel}

\begin{figure}
\centering
\includegraphics[width=1.00\linewidth]{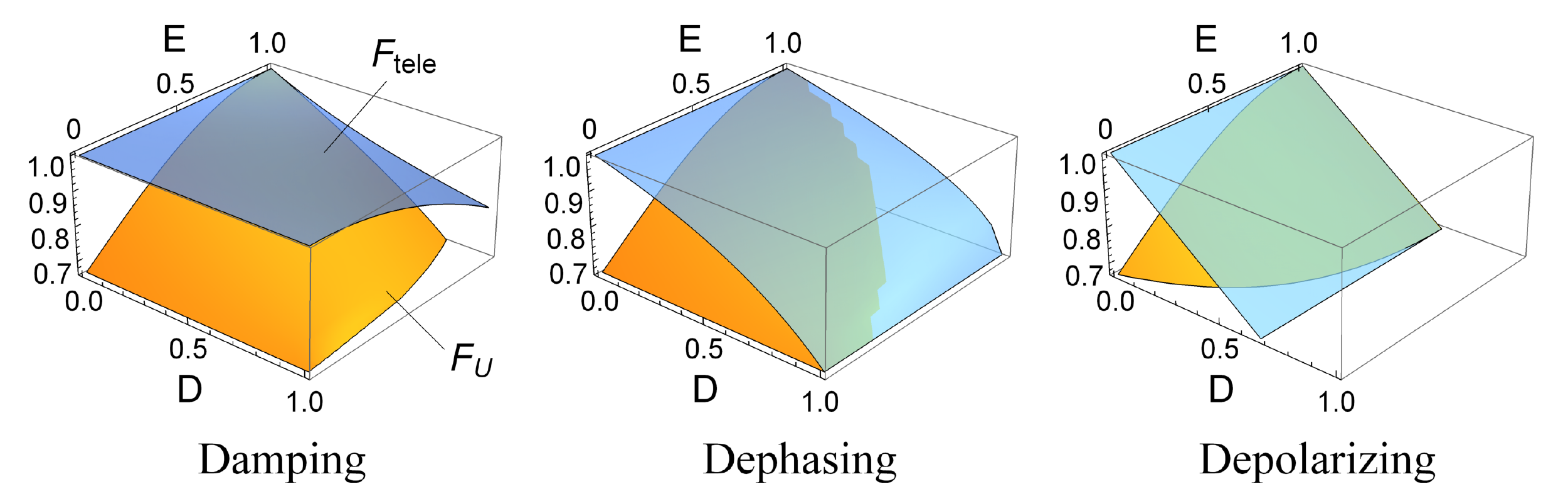}
\caption{Teleportation fidelities of the protocol optimized in MR framework under the effect of noise. Fidelities over the classical limit (${\cal F}_{\rm tele} > {\cal F}_{\rm cl}=2/3$) are plotted by changing the initial entanglement degree $E$ and the noise strength $D$. For comparison, the teleportation fidelities ${\cal F}_{U}$ of the conventional protocol based on unitary operations also are plotted.}
\label{fig:fig2}
\end{figure}

We now show that teleportation in MR framework can overcome the limit of conventional teleportation over noisy quantum channels. An arbitrary noise affecting the quantum channel can be described by a set of operators ${\cal E}=\{\hat{E}_k\}$, the explicit form of which varies by unitary recombination ($\cal U$) \cite{QIQM}. The corresponding coarse-grained  measurement ${\bf M}=\{\hat{M}_{i,k}\}$ can be defined by (\ref{eq:efmn}). For each outcome $i$, we can consider all corresponding measurement operators $\{\hat{M}_{i,k=1},\hat{M}_{i,k=2},\ldots \}$ and find the maximum among their smallest singular values $\{\lambda_{\rm min}^{i,k=1},\lambda_{\rm min}^{i,k=2},\ldots \}$ as $\lambda^{i,k_m}_{\rm min}\equiv \max_{k, {\cal U}}{[\lambda_{\rm min}^{i,k}}]$.
We then define a reversing operator as 
\begin{equation}
\label{eq:oRn}
\hat{R}^i=\lambda^{i,k_m}_{\rm min}\hat{U}^{\dag}_{i,k_m}\hat{D}^{-1}_{i,k_m}\hat{V}^{\dag}_{i,k_m}.
\end{equation}
As a final step, Bob's optimal reversing operation is chosen between a conditional unitary operation in a conventional scheme and the reversing operator (\ref{eq:oRn}) so as to yield the maximum teleportation fidelity
\begin{equation}
{\cal F}_{\rm tele}=\max[{\cal F}_U, {\cal F}_R].
\end{equation}

As paradigmatic examples, we consider damping, dephasing and depolarization in quantum channel $\ket{\Psi}=\cos{(\theta/2)}\ket{00}+\sin{(\theta/2)}\ket{11}$ (See Appendix~\ref{asec:DeProto} for details). In Fig.~\ref{fig:fig2}, we plot the maximum teleportation fidelities optimized in MR framework against the entanglement degree of quantum channel $E=\sin(\theta/2)$ and the noise strength $D$. The result shows that the optimal teleportation in MR framework overcomes the effect of noise beyond the reach of conventional teleportation ${\cal F}_{\rm tele} \geq {\cal F}_U$. In particular, we observe that, in some region of $D$ and $E$ where ${\cal F}_U<2/3$, the teleportation fidelities in MR framework surpass the classical limit ${\cal F}_{\rm tele}>{\cal F}_{\rm cl}=2/3$. It is a clear evidence that quantum teleportation is possible even under such a noisy quantum channel for which  teleportation was deemed impossible to date. Moreover, this is a teleportation using a single-copy of entangled state of the quantum channel without necessitating additional resources and entanglement distillation \cite{Bennett97,pan03,yama03}. An experimental demonstration of the optimized protocols will be presented elsewhere \cite{Im20}. 

\section{Applications}

Let us apply MR framework to optimize teleportation-based functionalities potentially useful in building quantum processors and networks.

\subsection{Multipartite teleportation}

Alice teleports a qubit $\ket{\psi}_{a}$ to Charlie by cooperating with Bob via a GHZ-type entangled channel $\ket{\Psi}_{\bar{a}bc}=\cos(\theta/2)\ket{000}+\sin(\theta/2)\ket{111}$ ($0 \leq \theta \leq \pi/2$) (see Fig.~\ref{fig:App}(a)) \cite{Karlsson98,Hillery99,Zhao04,Yonezawa04,Barasi19}. Alice performs the Bell-state measurement on $\ket{\psi}_{a}$ and her part of the channel. Bob performs a local measurement on his part. Then, they send their outcomes $i$ and $i'$ to Charlie. In this scenario, the maximum teleportation fidelity based on unitary reversal is obtained as ${\cal F}_U=(2+\sin{\theta}\sin{\phi})/3$. In MR framework, the corresponding measurement operator is $\hat{M}_{i,i'}\equiv {}_b\bra{w_{i'}}{}_{a\bar{a}}\bra{v_{i}}\cdot \ket{\Psi}_{\bar{a}bc}$, where $\ket{v_{i}}$ denotes the Bell basis and $\ket{w_1}=\cos(\phi/2)\ket{0}+\sin(\phi/2)\ket{1}$ and $\ket{w_2}=-\sin(\phi/2)\ket{0}+\cos(\phi/2)\ket{1}$ ($0 \leq \phi \leq \pi/2$) are Bob's measurement basis. We can find $\hat{R}^{i,i'}$ for each $\hat{M}_{i,i'}$ such that $\hat{R}^{i,i'}\hat{M}_{i,i'}\ket{\psi}\propto\ket{\psi}$. For instance, when $i=1$ and $i'=1$, $\hat{M}_{i=1,i'=1}=(1/\sqrt{2})(\cos(\theta/2)\cos(\phi/2)\ket{0}\bra{0}+\sin(\theta/2)\sin(\phi/2)\ket{1}\bra{1})$. Its optimal reversing operator is $R^{i=1,i'=1}=\tan(\theta/2)\tan(\phi/2)\ket{0}\bra{0}+\ket{1}\bra{1}$. By this, a faithful tripartite teleportation ${\cal F}_{\rm tele}=1$ is possible with success probability ${\cal P}^{\rm max}_{\rm tele}=2\sin^2(\min[\theta,\phi]/2)$. The amount of information leaked to Alice and Bob is obtained as ${\cal G}^{\rm max}=(2-\sin^2(\min[\theta,\phi]/2))/3$.
These saturate the upper bound of the trade-off relation (\ref{eq:balance}). Note that the analysis above can generally be extended to arbitrary multipartite teleportation.

\subsection{Entanglement transmission}
Suppose that Alice transfers a two-qubit state $\ket{\psi}_{ac}=\alpha\ket{00}+\beta\ket{10}+\gamma\ket{01}+\delta\ket{11}$ to Bob via two entangled pairs $\ket{\Psi}_{\bar{a}b}=\cos(\theta/2)\ket{00}+\sin(\theta/2)\ket{11}$ and $\ket{\Psi}_{\bar{c}d}=(\ket{00}+\ket{11})/\sqrt{2}$ (see Fig.~\ref{fig:App}(b)). The Bell-state measurement is performed between $a$ and $\bar{a}$, while the joint measurement between $c$ and $\bar{c}$ is performed in $\ket{v'_1}=\cos(\phi/2)\ket{00}+\sin(\phi/2)\ket{11}$, $\ket{v'_2}=\sin(\phi/2)\ket{00}-\cos(\phi/2)\ket{11}$, $\ket{v_3}=\cos(\phi/2)\ket{01}+\sin(\phi/2)\ket{10}$, $\ket{v'_4}=\sin(\phi/2)\ket{01}-\cos(\phi/2)\ket{10}$. 
In this scenario, a protocol based on unitary reversal \cite{JLee00,Zhang06} yields at best ${\cal F}_U=(1+(1+\sin\theta)(1+\sin\phi))$.
In MR framework, e.g., for the outcomes $i=1$ and $i'=2$, $\hat{M}_{i=1,i'=2}=(1/2)(\cos(\theta/2)\ket{0}_{ba}\bra{0}+\sin(\theta/2)\ket{1}_{ba}\bra{1})\otimes (\sin(\phi/2)\ket{0}_{dc}\bra{0}-\cos(\phi/2)\ket{1}_{dc}\bra{1})$. Its optimal reversing operator is then $R^{i=1,i'=2}=(\tan(\theta/2)\ket{0}\bra{0}+\ket{1}\bra{1})\otimes(-\tan(\phi/2)\ket{1}\bra{1}+\ket{0}\bra{0})$. As a result, a transmission of arbitrary entangled qubits is possible with ${\cal F}_{\rm tele}=1$ and success probability ${\cal P}^{\rm max}_{\rm tele}=4\sin^2(\theta/2)\sin^2(\phi/2)$. This is also applicable for the transfer of arbitrary multipartite entanglement. See Appendix~\ref{asec:app} for details.

\subsection{One-way quantum repeater}
To transfer qubits over long distance, quantum repeaters are required \cite{duan01}. One-way repeater exploits a teleportation-based scheme to relay qubits at intermediate nodes \cite{Munro12,Muralidharan12,Lee19}. Assume that Alice transmits a qubit $\ket{\psi}_a$ over long distance to David (see Fig.~\ref{fig:App}(c)). To relay the qubit, Bob performs the Bell-state measurement between the qubit from Alice and one of the entangled pair $\ket{\Psi}_{bc}=\cos(\theta/2)\ket{00}+\sin(\theta/2)\ket{11}$ and transmits the remaining qubit to Charlie. Similarly, Charlie relays the qubit to David with maximally entangled qubits prepared for the channel and the joint measurement taken as $\ket{v'_1}=\cos(\phi/2)\ket{00}+\sin(\phi/2)\ket{11}$, $\ket{v'_2}=\sin(\phi/2)\ket{00}-\cos(\phi/2)\ket{11}$, $\ket{v'_3}=\cos(\phi/2)\ket{01}+\sin(\phi/2)\ket{10}$, $\ket{v'_4}=\sin(\phi/2)\ket{01}-\cos(\phi/2)\ket{10}$.

In this scenario, the maximum attainable fidelity by unitary reversal is ${\cal F}_U=(2+\sin{\theta}\sin{\phi})/3$. In contrast, MR framework yields an optimal operator for David to faithfully recover the input qubit. For example, when Bob's and Charlie's outcomes are $i=2$ and $i'=1$, respectively, $\hat{M}_{i=2,i'=1}=(1/2)(\cos(\theta/2)\cos(\phi/2)\ket{0}\bra{0}-\sin(\theta/2)\sin(\phi/2))$. David's optimal operator is then $\hat{R}^{i=2,i'=1}=\tan(\theta/2)\tan(\phi/2)\ket{0}\bra{0}-\ket{1}\bra{1}$ such that $\hat{R}^{i=2,i'=1}\hat{M}_{i=2,i'=1}\ket{\psi}\propto\ket{\psi}$. As a result, a faithful long-distance  transmission ${\cal F}_{\rm tele}=1$ is possible with success probability ${\cal P}^{\rm max}_{\rm tele}=2\sin^2(\min[\theta,\phi]/2)$ (details in Appendix~\ref{asec:app}).

\begin{figure}
\centering
\includegraphics[width=0.99\linewidth]{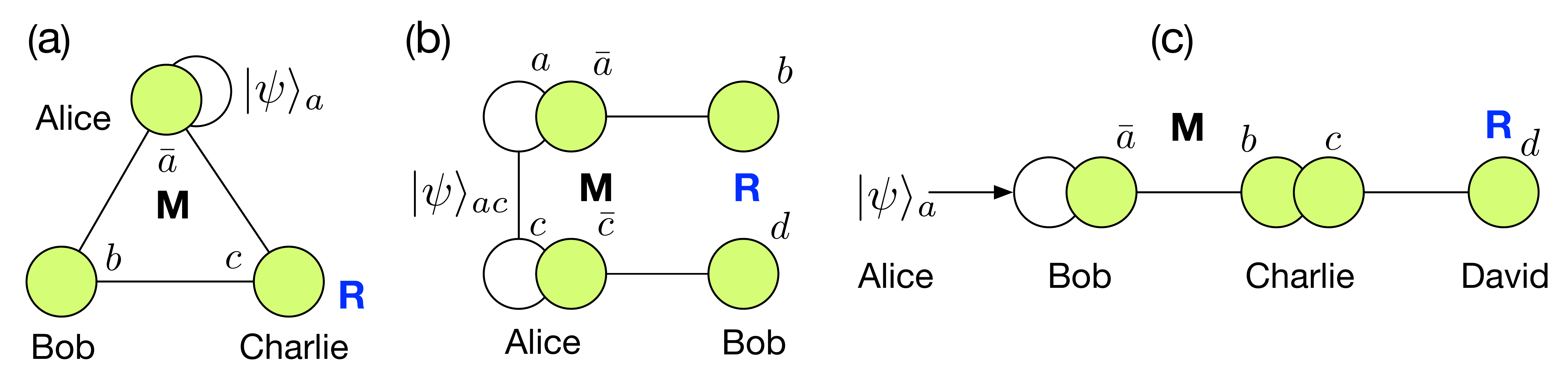}
\caption{Applications for (a) multipartite teleportation, (b) entanglement transmission, (c) one-way quantum repeater. MR framework makes it possible to optimize these protocols to transfer an unknown quantum state $\ket{\psi}$ with the unit fidelity ${\cal F}_{\rm tele}=1$ and maximum success probability ${\cal P}^{\rm max}_{\rm tele}$.}
\label{fig:App}
\end{figure}

\section{Remarks}

Our result can be directly used to implement gate operations for near-term noisy quantum processors where the number of available quantum resources would be limited. In MR framework, we can find an optimal reversing operation for the teleportation over an arbitrary noisy quantum channel $\ket{\Psi}_{\bar{a}b}$. This protocol is also optimal for the teleportation over the channel $(\hat{I}\otimes\hat{G})\ket{\Psi}_{\bar{a}b}$ to yield the output state $\hat{G}\ket{\psi}$. Thus, following \cite{Gottesman99}, we can implement an arbitrary gate operation $\hat{G}$ under the influence of noise without requiring additional qubits.

A recent work \cite{Cavalcanti17} reported that any entangled state yields non-classical teleportation in the sense that the assemblage of the conditional states upon Alice's measurement cannot be reproduced by unentangled states. 
In our work, we take a different perspective aiming at high fidelity over classical teleportation limit for a given entangled state under noise. Our MR framework enables us to achieve the maximum teleportation fidelity with the maximum success probability up to the fundamental limit, extended to arbitrary-dimensional multipartite teleportation. It also offers a realistic way to realize teleportation via a noisy quantum channel overcoming the limit of previous protocols. Moreover, our protocol is easily implementable as it requires only single-qubit operation at receiver's party, while previous proposals for optimal teleportation \cite{Banaszek00,Mor99,WSon00, Roa03,Li00,Agrawal02} rely on operations difficult to realize, e.g. arbitrary two-qubit POVMs. 
We hope that our proposed MR framework would provide a further insight to mitigate the effect of noise while transmitting quantum information over noisy quantum channel stimulating related works. 

\acknowledgments

This work was supported in part by the National Research Foundation of Korea (2019R1A2C3004812, 2020M3E4A1079939), the KIST institutional program (2E31021), the ITRC support program (IITP-2020-0-01606), Agency for Defense Development (UG190028RD), the KIST Open Research Program, the QuantERA ERA-NET within the EU Horizon 2020 Programme, and the EPSRC (EP/R044082/1). 

\appendix
\section{Measurement-Reversal framework for teleportation}

\subsection{Effective quantum measurement for teleportation}
\label{asec:eqm}

Let us consider a general scenario of one-to-one quantum teleportation. An unknown quantum state $\rho_a$ is prepared in mode $a$ and an entangled quantum state $\Psi_{\bar{a}b}$ is distributed between two modes $\bar{a}$ (at Alice's location) and $b$ (at Bob's location). Alice performs an arbitrary joint measurement $\{\Pi^i\}$ between mode $a$ and $\bar{a}$. Then, the unnormalized reduced state is given by
\begin{equation}
\label{eq:stateb}
\tilde{\rho}^i_b=\Pi_{a\bar{a}}^i(\rho_a\otimes\Psi_{\bar{a}b})=M^i_{a\rightarrow b}(\rho).
\end{equation}
Therefore, we can define an effective quantum measurement $\{M_{a\rightarrow b}^i\}$ that is nonlocal in the sense that its input($a$) and output($b$) modes are spatially separated. 
The joint measurement can be generally defined by operators $\hat{\Pi}^i_{a\bar{a}}=\sum_l \delta^{(i)}_l\ket{v^{(i)}_l}\bra{w^{(i)}_l}$ where $\{\ket{v^{(i)}_l}\}$ and $\{\ket{w^{(i)}_l}\}$ are arbitrary orthonormal basis in the Hilbert space of ${\cal H}_a \otimes {\cal H}_{\bar{a}}$, $\delta_l$ are non-negative values and $\sum_i \hat{\Pi}^{i \dag}_{a\bar{a}}\hat{\Pi}^{i}_{a\bar{a}}=1$. An arbitrary entangled state can be given in the form of $\Psi_{\bar{a}b}=\sum_kp_k\ket{\Psi_k}\bra{\Psi_k}$. We then define its corresponding non-local quantum measurement with operators 
\begin{equation}
\hat{M}_{i,k,l}\equiv\sqrt{p_k}\delta^{(i)}_l {}_{\bar{a}b}\bracket{\Psi_k}{w_l^{(i)}}_{a\bar{a}},
\end{equation}
and $\sum_{i,k,l} \hat{M}^\dag_{i,k,l}\hat{M}_{i,k,l}=\sum_{i,k,l} p_k(\delta^{(i)}_l)^2 \bracket{\Psi_k}{w_l^{(i)}}_{a\bar{a}} \bracket{w_l^{(i)}}{\Psi_k}=1$, by which the reduced state is then represented by 
\begin{equation}
\tilde{\rho}^i_b={\rm tr}_{a\bar{a}}[\Pi^i_{a\bar{a}} (\rho_a \otimes \Psi_{\bar{a}b})]=\sum_{k,l} \hat{M}_{i,k,l} \rho_a \hat{M}_{i,k,l}^{\dag}.
\end{equation}
Therefore, once the outcome $i$ is shared with Bob, the reduced state is equivalent to a post-measurement state obtained when a quantum measurement ${\bf M}=\{ \hat{M}_{i,k,l}\}$ is applied to an unknown quantum state $\rho$ at Bob's location.
If we assume a rank-one joint measurement $\Pi^i_{a\bar{a}}=\ket{v_i}\bra{v_i}$ and a pure entangled state $\Psi_{\bar{a}b}=\ket{\Psi}\bra{\Psi}$, the reduced state is
\begin{equation}
\label{eq:effm}
\tilde{\rho}^i_b={}_{a\bar{a}}\bra{v_{i}}\cdot\rho_a \otimes \ket{\Psi}_{\bar{a}b}\bra{\Psi}\cdot\ket{v_{i}}_{a\bar{a}}
=\hat{M}_i\rho\hat{M}_i^{\dag},
\end{equation}
where the effective non-local quantum measurement is defined by $\hat{M}_i={}_{a\bar{a}}\bracket{v_{i}}{\Psi}_{\bar{a}b}$ satisfying $\sum_i \hat{M}_i^{\dag}\hat{M}_i=1$. 

\subsection{Measurement-Reversal framework and performance measures}
\label{asec:pm}

An arbitrary teleportation protocol can be in general described by $({\bf R}\circ{\bf M}) (\rho) \propto \tilde{\rho}$ in terms of a measurement ${\bf M}$ and a reversing operation ${\bf R}$ optimized toward $\tilde{\rho}=\rho$. First, assume a pure input state $\rho=\ket{\psi}\bra{\psi}$ and an ideal quantum measurement  ${\bf M}=\{\hat{M}_i\}$. For the measurement outcome $i$, the post-measurement state is given by $\rho^i=\hat{M}_i \ket{\psi}\bra{\psi}\hat{M}_i^{\dag}/p^i(\psi)$ where $p^i(\psi)= \bra{\psi}\hat{M}_i^{\dag}\hat{M}_i \ket{\psi}$. The teleported state after applying ${\bf R}$ (obtaining its outcome $j$) is then written by $\rho^i_j=\hat{R}^i_{j}\hat{M}_{i} \ket{\psi}\bra{\psi}\hat{M}_{i}^{\dag}\hat{R}_{j}^{i \dag}/p^i_j(\psi)$, where $p^i_j(\psi)=\sum \bra{\psi}\hat{M}_{i}^{\dag}\hat{R}_{j}^{i \dag}\hat{R}^i_{j}\hat{M}_{i} \ket{\psi}$. Note that each quantum measurement operator can be represented in the singular value decomposition as $\hat{M}_{i}=\hat{V}_i\hat{D}_i\hat{U}_i$ with unitary operators $\hat{V}_i$ and $\hat{U}_i$ and a diagonal matrix $\hat{D}_i=\sum_{n=0}^{d-1}\lambda^i_n\ket{n}\bra{n}$ in arbitrary $d$-dimensional Hilbert space. We can then find an optimal reversing operator as $\hat{R}^i=\lambda^i_{\rm min}\hat{U}^\dag_i\hat{D}^{-1}_i\hat{V}^\dag_i$ so that an arbitrary input state can be faithfully recovered by $\hat{R}^i\hat{M}_i\ket{\psi}=\lambda^i_{\rm min}\ket{\psi}$. 

This framework is also valid for a non-ideal quantum measurement and noisy quantum channel ${\cal E}=\{\hat{E}^{k}\}$. In such a case, the effective non-local quantum measurement is generally given as a coarse-grained measurement e.g.~${\bf M}=\{\hat{M}_{i,k}\}$ in the sense that it yields outcome $i$ while the outcome $k$ is hidden.
The post-measurement state is then given by $\rho^i=\sum _k \hat{M}_{i,k} \ket{\psi}\bra{\psi}\hat{M}^{\dag}_{i,k}/p^i(\psi)$ where $p^i(\psi)=\sum_k \bra{\psi}\hat{M}^{\dag}_{i,k}\hat{M}_{i,k} \ket{\psi}$. The teleported state then reads $\rho^i_j=\sum _k \hat{R}^i_{j}\hat{M}_{i,k} \ket{\psi}\bra{\psi}\hat{M}_{i,k}^{\dag}\hat{R}_{j}^{i \dag}/p^i_j(\psi)$ 
where $p^i_j(\psi)=\sum_k \bra{\psi}\hat{M}_{i,k}^{\dag}\hat{R}_{j}^{i \dag}\hat{R}^i_{j}\hat{M}_{i,k} \ket{\psi}$. 
Note that the noise operators can have different forms by changing the representation by unitary recombination ($\cal U$), and so do the measurement operators. Considering all the measurement operators for a single identified outcome $i$, i.e., $\{\hat{M}_{i,k=1},\hat{M}_{i,k=2},\ldots \}$, we find the maximum value  $\lambda^{i,k_m}_{\rm min}\equiv \max_{k, {\cal U}}{[\lambda_{\rm min}^{i,k}}]$ among the smallest singular values $\{\lambda_{\rm min}^{i,k=1},\lambda_{\rm min}^{i,k=2},\ldots \}$ over all possible representations $\cal U$, where $k=k_m$ gives the maximum $\lambda^{i,k_m}$.
We then define a reversing operator for each outcome $i$ as $\hat{R}^i=\lambda^{i,k_m}_{\rm min}\hat{U}^{\dag}_{i,k_m}\hat{D}^{-1}_{i,k_m}$.

The performance of quantum teleportation can be assessed in terms of the teleportation fidelity, the success probability and the information gain by Alice (senders and intermediators in multipartite protocols) as listed below:

\begin{itemize}

\item[(i)] {\em Teleportation fidelity}-- We can define the teleportation fidelity as the average overlap between the input and the output state obtained when the teleportation succeeds, i.e.,
\begin{equation}
\label{eq:Ftele}
{\cal F}_{\rm tele}=\int d \psi \sum_{i} p^i(\psi) \bra{\psi} \rho^i_{j=1} \ket{\psi}.
\end{equation}
For the teleportation via a pure entangled state without noise, a faithful teleportation is always possible. i.e., ${\cal F}_{\rm tele}=1$, by applying an optimal reversing operator for a given ${\bf M}=\{\hat{M}_i\}$. On the other hand, in the presence of noise, the input pure state $\ket{\psi}$ cannot be faithfully recovered. The attainable teleportation fidelity is thus lower than the unity $F<1$. Bob's optimal reversing operation is determined to yield the maximum teleportation fidelity, by comparing the fidelity obtained by the conditional unitary operation ${\cal F}_U$ in the conventional scheme and the fidelity obtained by the reversing operation ${\cal F}_R$, i.e., ${\cal F}_{\rm tele}=\max[{\cal F}_U, {\cal F}_R]$ in MR framework.

\item[(ii)] {\em Success probability}-- We define the average success probability of the teleportation as
\begin{equation}
\label{eq:Ptele}
{\cal P}_{\rm tele} = \int d \psi \sum_{i} p^i_{j=1}(\psi).
\end{equation}
where $p^i_{j=1}$ is the success probability of the reversing operation for the measurement outcome $i$. We here assume that each reversing operation succeeds when its outcome is $j=1$ without loss of generality. In the absence of noise in quantum channel, the maximum success probability can be obtained as ${\cal P}^{\rm max}_{\rm tele} =\sum_i (\lambda^i_{\rm min})^2$, where $\lambda^i_{\rm min}$ is the smallest singular value of $\hat{M}_i$. Detailed methods appear in Ref.~\cite{Cheong12}.

\item[(iii)] {\em Information gain by Alice}-- The amount of information extracted by Alice (or senders and intermediators in multipartite scenarios) affects the teleportation performance. Assume that Alice attempts to extract information of the input state $\ket{\psi}$ and estimate it as $\ket{\widetilde{\psi}_i}$ for the outcome $i$ based on the knowledge of ${\bf M}$. Then, the information gain obtained by Alice can be quantified by averaging the closeness between $\ket{\psi}$ and $\ket{\tilde{\psi}_i}$,
\begin{equation}
{\cal G}_{\rm Alice}= \int d\psi \sum_{i}\sum_k p^{i,k}(\psi) |\bracket{\tilde{\psi}_i}{\psi}|^2,
\end{equation}
where $p^{i,k}(\psi)= \bra{\psi}\hat{M}_{i,k}^{\dag}\hat{M}_{i,k} \ket{\psi}$. Without the effect of noise in quantum channel, the maximum gain can be evaluated as ${\cal G}^{\rm max}_{\rm Alice}=(d+\sum_i(\lambda^i_{\rm max})^2)/(d+1)$, where $\lambda^i_{\rm max}$ is the largest singular value of $\hat{M}_i$  \cite{Bana}.

\end{itemize}

\section{Fundamental limit by the no-cloning theorem} 
\label{asec:FL}

\begin{figure}[b]
\centering
\includegraphics[width=0.8\linewidth]{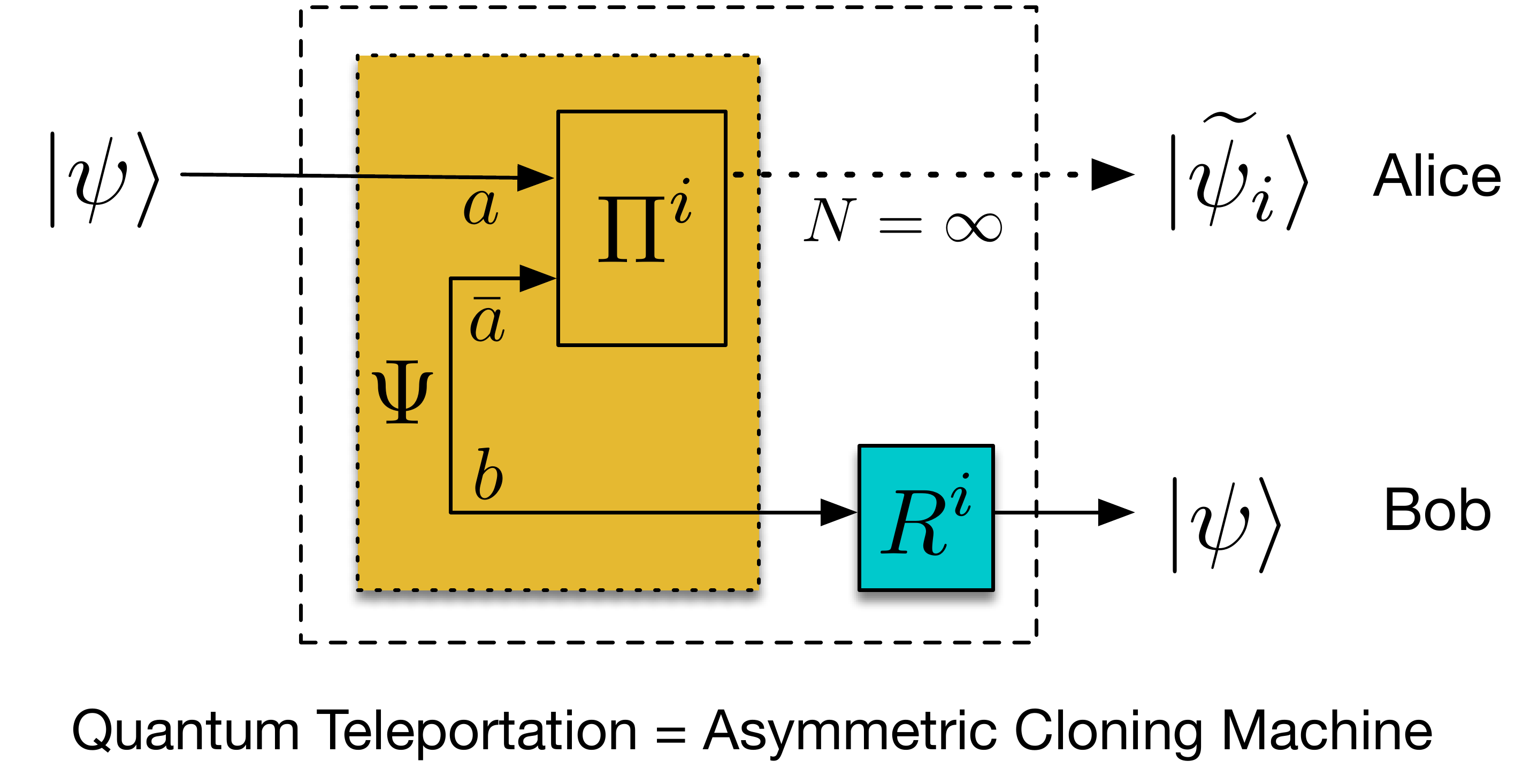}
\caption{\textbf{Teleportation is an asymmetric cloning.} A faithful teleportation is $1\rightarrow N+1$ asymmetric cloning with $N \rightarrow \infty$.
}
\label{figTeleclon}
\end{figure}

The performance of quantum teleportation is fundamentally restricted by the no-cloning theorem. This can be understood in a heuristic way as any leakage of information about an unknown input state $\ket{\psi}$ during the teleportation process would decrease the maximum amount of information that can be transmitted to Bob and the overall performance is degraded. MR framework allows us to find such fundamental limits of the teleportation performance in an exact quantitative manner. As illustrated in Fig.~\ref{figTeleclon}, any faithful quantum teleportation process can be regarded as a $1\rightarrow N+1$ asymmetric cloning, i.e., $\ket{\psi}\rightarrow \ket{\tilde{\psi}_i}^{\otimes N}\ket{\psi}$ with $N \rightarrow \infty$. The upper output of the cloning process yields an infinite number of copies of the estimated state by Alice, and its output fidelity is evaluated as the information gain by Alice ${\cal G}^{\rm max}_{\rm Alice}$. On the other hand, the bottom output yields the teleported state that is exactly the same with the input state. Its output fidelity is thus given by the teleportation fidelity ${\cal F}_{\rm tele}=1$, and the overall maximum success probability is given by ${\cal P}^{\rm max}_{\rm tele}$. The performance of this cloning machine is restricted by a trade-off relation between the information gain by Alice ${\cal G}^{\rm max}_{\rm Alice}$ and the success probability ${\cal P}^{\rm max}_{\rm tele}$ in Eq.~(\ref{eq:balance}). This generally holds for any teleportation scenarios including arbitrary multiparty and $d$-dimensional teleportations.

\section{Comparison with other protocols for faithful teleportation} 
\label{asec:Comp}

MR framework provides a way to optimize teleportation protocol to yield the maximum success probability ${\cal P}^{\rm max}_{\rm tele}$ saturating the upper bound of Eq.~(\ref{eq:balance}), while maintaining the teleportation fidelity ${\cal F}_{\rm tele}=1$. On the other hand, no previous protocols for faithful teleportation such as conclusive \cite{Mor99,WSon00,Roa03} and probabilistic teleportation \cite{Li00,Agrawal02} address such a fundamental limit of performance. It is also nontrivial to optimize teleportation protocol for a given arbitrary quantum channel up to the fundamental limit in previous protocols. This is because the optimization of conclusive or probabilistic teleportations is only based on the modification of Alice's joint measurement, while MR framework generally allows us to modify not only the joint measurement but also Bob's reversing operation. 

Let us consider a qubit teleportation scenario for comparison. Assume that Alice transmits an unknown qubit $\ket{\psi}_{a}$ via a quantum channel prepared in $\ket{\Psi}_{\bar{a}b} = \cos{\frac{\theta}{2}}\ket{0}_{\bar{a}}\ket{0}_b+\sin{\frac{\theta}{2}}\ket{1}_{\bar{a}}\ket{1}_b$ where $0 \leq \theta \leq \frac{\pi}{2}$, which becomes a maximally entangled state or a product state by setting $\theta=\frac{\pi}{2}$ or $\theta=0$, respectively. Alice performs a joint measurement on the basis $\{ \ket{W_1}_{a\bar{a}}=\cos{\frac{\phi}{2}}\ket{0}_{a}\ket{0}_{\bar{a}}+\sin{\frac{\phi}{2}}\ket{1}_a\ket{1}_{\bar{a}}, \ket{W_2}_{a\bar{a}}=\sin{\frac{\phi}{2}}\ket{0}_a\ket{0}_{\bar{a}}-\cos{\frac{\phi}{2}}\ket{1}_a\ket{1}_{\bar{a}}, \ket{W_3}_{a\bar{a}}=\cos{\frac{\phi}{2}}\ket{0}_a\ket{1}_{\bar{a}}+\sin{\frac{\phi}{2}}\ket{1}_a\ket{0}_{\bar{a}}, \ket{W_4}_{a\bar{a}}=\sin{\frac{\phi}{2}}\ket{0}_a\ket{1}_{\bar{a}}-\cos{\frac{\phi}{2}}\ket{1}_a\ket{0}_{\bar{a}} \}$, which becomes the Bell basis when $\phi=\frac{\pi}{2}$. In MR framework, the quantum measurement representation of the teleportation is described by a set of operators $\{ \hat{M}_i = {}_{a\bar{a}}\bra{W_i}\ket{\Psi}_{\bar{a}b} \}$ with 
\begin{equation}
\begin{aligned}
\label{eq:nonlocalm}
\hat{M}_1=\cos{\frac{\theta}{2}}\cos{\frac{\phi}{2}}\ket{0}_b~{}_{a}\bra{0}+\sin{\frac{\theta}{2}}\sin{\frac{\phi}{2}}\ket{1}_b~{}_{a}\bra{1}\\
\hat{M}_2=\cos{\frac{\theta}{2}}\sin{\frac{\phi}{2}}\ket{0}_b~{}_{a}\bra{0}-\sin{\frac{\theta}{2}}\cos{\frac{\phi}{2}}\ket{1}_b~{}_{a}\bra{1}\\
\hat{M}_3=\cos{\frac{\theta}{2}}\cos{\frac{\phi}{2}}\ket{0}_b~{}_{a}\bra{1}+\sin{\frac{\theta}{2}}\sin{\frac{\phi}{2}}\ket{1}_b~{}_{a}\bra{0}\\
\hat{M}_4=\cos{\frac{\theta}{2}}\sin{\frac{\phi}{2}}\ket{0}_b~{}_{a}\bra{1}-\sin{\frac{\theta}{2}}\cos{\frac{\phi}{2}}\ket{1}_b~{}_{a}\bra{0}
\end{aligned}
\end{equation}
which satisfies $\hat{M_1}^{\dag}\hat{M_1}+\hat{M_2}^{\dag}\hat{M_2}+\hat{M_3}^{\dag}\hat{M_3}+\hat{M_4}^{\dag}\hat{M_4}=1$. 
The optimal reversing operators for each measurement outcome $i=1,2,3,4$ are then given as
\begin{equation}
\label{eq:ORop}
\begin{aligned}
\hat{R}^{1}&=\tan{\frac{\theta}{2}}\tan{\frac{\phi}{2}}\ket{0}\bra{0}+\ket{1}\bra{1},\\
\hat{R}^{2}&=\begin{cases}
\hat{\sigma}_x\Big(\cot{\frac{\theta}{2}}\tan{\frac{\phi}{2}}\ket{0}\bra{0}+\ket{1}\bra{1}\Big)i\hat{\sigma}_y, & \text{if $\theta \geq \phi$},\\
\hat{\sigma}_x\Big(\ket{0}\bra{0}+\tan{\frac{\theta}{2}}\cot{\frac{\phi}{2}}\ket{1}\bra{1}\Big)i\hat{\sigma}_y, & \text{if $\theta < \phi$}
\end{cases}\\
\hat{R^{3}}&=\hat{\sigma}_x\Big(\tan{\frac{\theta}{2}}\tan{\frac{\phi}{2}}\ket{0}\bra{0}+\ket{1}\bra{1}\Big)\\
\hat{R^{4}}&=\begin{cases}
\Big(\cot{\frac{\theta}{2}}\tan{\frac{\phi}{2}}\ket{0}\bra{0}+\ket{1}\bra{1}\Big)i\hat{\sigma}_y, & \text{if $\theta \geq \phi$},\\
\Big(\ket{0}\bra{0}+\tan{\frac{\theta}{2}}\cot{\frac{\phi}{2}}\ket{1}\bra{1}\Big)i\hat{\sigma}_y, & \text{if $\theta < \phi$}.
\end{cases}
\end{aligned}
\end{equation}
After receiving the measurement outcome $i$ through a classical channel from Alice, Bob applies an appropriate reversing operation $\{B^i_{j}\}$ in Eq.~(\ref{eq:ORop}). The input state can then be faithfully recovered at Bob's party when Bob's reversing operation succeeds (i.e., when the outcome is $j=1$). The overall success probability of the teleportation protocol can then be evaluated as 
\begin{equation}
\label{eq:Ptele}
{\cal P}_{\rm tele}=2\sin^2\Big(\frac{{\rm min}[\theta,\phi]}{2}\Big).
\end{equation}
For a given quantum channel in a non-maximally entangled state ($\theta<\pi/2$), a faithful teleportation can be achieved with the success probability ${\cal P}^{\rm max}_{\rm tele}=2\sin^2(\theta/2)$ for $\phi > \theta$. It shows a monotonic increase of the success probability of quantum teleportation as the degree of entanglement of the channel $E=\sin{(\theta/2)}$ increases. The maximum amount of information gain by Alice can be also evaluated as ${\cal G}^{\rm max}_{\rm Alice}=(2-\sin^{2}(\theta/2))/3$. Note that the success probability of the teleportation in Eq.~(\ref{eq:Ptele}) reaches the upper bound in Eq.~(\ref{eq:balance}) together with the information gain ${\cal G}^{\rm max}_{\rm Alice}$. 

First, let us consider a probabilistic teleportation protocol proposed in Ref.~\cite{Agrawal02} for comparison, which uses quantum channel, $\ket{\Psi}_{\bar{a}b}=(\ket{00}+n\ket{11})/\sqrt{1+|n|^2}$ for $0\leq n \leq 1$. This allows the unit teleportation fidelity ${\cal F}_{\rm tele}=1$ by modifying the joint measurement basis as $\ket{W_1}=(\ket{00}+n\ket{11})/\sqrt{1+|n|^2}$, $\ket{W_2}=(n^{*}\ket{00}-\ket{11})/\sqrt{1+|n|^2}$, $\ket{W_3}=(n\ket{01}+\ket{10})/\sqrt{1+|n|^2}$, $\ket{W_4}=(\ket{01}-n^{*}\ket{10})/\sqrt{1+|n|^2}$ and post-selecting the output qubits when the measurement outcome is $i=2$ or 4. The success probability of this protocol is ${\cal P}_{\rm tele}=2|n|^2/(1+|n|^2)^2$. However, in MR framework, the protocol can be further optimized by additionally applying reversing operation at Bob's party even when the outcome is $i=1$ or $3$, i.e., $\hat{R}^1=|n|^2\ket{0}\bra{0}+\ket{1}\bra{1}$ and $\hat{R}^3=|n|^2\ket{1}\bra{0}+\ket{0}\bra{1}$, which also yield events in which the unknown input qubit $\ket{\psi}$ is perfectly teleported. This example illustrates that the protocol in~\cite{Agrawal02} does not fully use the quantum resources. As a result, MR framework allows a faithful teleportation ${\cal F}_{\rm tele}=1$ with the success probability ${\cal P}^{\rm max}_{\rm tele}=2|n|^2/(1+|n|^2)$ that is always higher than ${\cal P}_{\rm tele}=2|n|^2/(1+|n|^2)^2$. The maximum amount of information that Alice can extract in this case is ${\cal G}^{\rm max}_{\rm Alice}=(2+3|n|^2+|n|^4)/3(1+|n|^2)^2$. It is straightforward to see that the obtained success probability ${\cal P}^{\rm max}_{\rm tele}$ and the information gain ${\cal G}^{\rm max}_{\rm Alice}$ saturate the upper bound in Eq.~(\ref{eq:balance}).

Now, let us consider a conclusive teleportation protocol by following the method in Ref.~\cite{WSon00}. For a given channel $\ket{\Psi}_{\bar{a}b}=\sum^{d-1}_{\alpha=0}a_{\alpha}\ket{\alpha,\alpha}$, a conclusive teleportation can be optimized by modifying the joint measurement performed by Alice. The measurement operators for conclusive events can be defined as a rank one projector $\hat{M}_{i}=\lambda\ket{W_i}\bra{W_i}$ where $\ket{W_i} \equiv (1/d)\sum^{d-1}_{\alpha,\beta=0}(\hat{U}^{i}_{\alpha,\beta}a^{-1}_{\beta})\ket{\alpha,\beta}$ is the projection basis, and the measurement operator for inconclusive events is $\hat{M}_{d^2+1}=I-\sum^{d^2}_{i=1}\hat{M}_{i}$ from the completeness relation. We can rewrite the conclusive measurement operators as $\sum^{d^2}_{i=1}\hat{M}_{i}=\sum_{\alpha\beta}(\lambda/d a^2_\beta)\ket{\alpha,\beta}\bra{\alpha,\beta}$ by using the relation $\sum_i\hat{U}^{i}_{\alpha\beta}\hat{U}^{* i}_{\gamma\delta}=d\delta_{\alpha\gamma}\delta_{\beta\delta}$. From the positivity condition of the inconclusive operator, i.e., $\bra{\psi}\hat{M}_{d^2+1}\ket{\psi}=\bra{\psi}\sum_{\alpha\beta}(1-\lambda/d a^2_\beta)\ket{\alpha,\beta}\bra{\alpha,\beta}\cdot\ket{\psi}\geq0,~\forall \ket{\psi}$, we can find that $0\leq \lambda \leq a^2_{\rm min} d$ where $a_{\rm min}$ is the smallest coefficient of the channel state. As the probability of each conclusive event is given by $\bra{W_i}\hat{M_i}\ket{W_i}/d^2=\lambda/d^2$, the overall success probability of the conclusive teleportation protocol is ${\cal P}_{\rm tele}=\sum^{d^2}_{i=1}\lambda/d^2=\lambda$ which has the maximum value $a^2_{\rm min} d$. Note that it allows us to reach the maximum success probability ${\cal P}^{\rm max}_{\rm tele}$ in some scenarios. For example, a qubit teleportation with a channel state $\ket{\Psi}_{\bar{a}b} = \cos{\frac{\theta}{2}}\ket{0}_{\bar{a}}\ket{0}_b+\sin{\frac{\theta}{2}}\ket{1}_{\bar{a}}\ket{1}_b$ can achieve the maximum success probability ${\cal P}^{\rm max}_{\rm tele}=2\sin^2(\theta/2)$ by optimally modifying the joint measurement performed by Alice, which is the same with maximum obtained in MR framework. However, in general, it is nontrivial to optimize a conclusive teleportation protocol to yield the maximum success probability ${\cal P}^{\rm max}_{\rm tele}$ because as modifying the joint measurement the amount of information gain by Alice also changes. In addition, even in a qubit teleportation, implementing a conclusive teleportation is demanding 
as it frequently requires to realize a nontrivial two-qubit POVMs, while only a single qubit operation suffices to complete the optimized protocol in MR framework.
Moreover, in contrast to probabilistic or conclusive protocols, MR framework is generally and straightforwardly extendable to arbitrary multipartite quantum teleportation and also to the teleportation over noisy quantum channel.

\begin{widetext}
\section{Details of the teleportation protocol via a noisy quantum channel}
\label{asec:DeProto}

In this section, we give more details on the optimized quantum teleportation via a noisy quantum channel in MR framework. We consider an unknown qubit  $\ket{\psi}_a$ and a quantum channel prepared in an entangled state  $\ket{\Psi}_{\bar{a}b} = \cos{\frac{\theta}{2}}\ket{0}_{\bar{a}}\ket{0}_b+\sin{\frac{\theta}{2}}\ket{1}_{\bar{a}}\ket{1}_b$ where $0 \leq \theta \leq \frac{\pi}{2}$, which becomes a maximally entangled state for $ \theta = \pi/2 $. We assume that the mode $\bar{a}$ or $b$ of the channel state $\ket{\Psi}_{\bar{a}b}$ experiences decoherence described by a set of operators $\{\hat{E}_{k,\bar{a}}\}$ or $\{\hat{E}_{k',b}\}$ in the Kraus representation, respectively. The quantum measurement is then represented by 
\begin{equation}
\hat{M}_{i,k}={}_{a\bar{a}}\bra{W_i}
\hat{E}_{k,\bar{a}}\hat{E}_{k',b}\ket{\Psi}_{\bar{a}b},
\end{equation} 
in the presence of noise on both modes.
We first consider optimal teleportation protocols for the amplitude damping noise either in mode $b$ (with Bob) or $\bar{a}$ (with Alice) as described below.

\begin{itemize}

\item[(i)] {\em Decoherence in mode $b$} -- Consider an example in which decoherence occurs in mode $b$. This is a realistic scenario if the entangled qubit pair is prepared by Alice and then one qubit of the pair is sent to Bob. The amplitude damping noise can be described by $ \hat{E}_{1} = |0\rangle \langle 0| +  \sqrt{1-\text{D}} |1\rangle \langle 1|$ and $\hat{E}_{2} = \sqrt{\text{D}} |0\rangle \langle 1| $ \cite{QIQM}. Let us fix the joint measurement basis as the Bell basis for simplicity. We then define the corresponding quantum measurement $ \hat{M}_{i,k} = {}_{a\bar{a}} \langle W_i |  \hat{E}_{k,b} | \Psi\rangle_{\bar{a}b} $ as 
\begin{equation}
\begin{aligned}
\hat{M}_{i=1,k=1} &= \frac{1}{\sqrt2} \cos \frac{\theta}{2} |0\rangle_b {}_{a} \langle0| + \sqrt{\frac{1-D}{2}} \sin \frac{\theta}{2} |1\rangle_b {}_{a} \langle1|, \,\,\,\,\,\quad \hat{M}_{i=1,k=2} = \sqrt{\frac{D}{2}} \sin \frac{\theta}{2} |0\rangle_b {}_{a} \langle1|, \\
\hat{M}_{i=2,k=1} &= \frac{1}{\sqrt2} \cos \frac{\theta}{2} |0\rangle_b {}_{a} \langle0| - \sqrt{\frac{1-D}{2}}  \sin \frac{\theta}{2} |1\rangle_b {}_{a} \langle1|, \,\,\,\,\,\quad \hat{M}_{i=2,k=2} = - \sqrt{\frac{D}{2}}  \sin \frac{\theta}{2} |0\rangle_b {}_{a} \langle1|, \\
\hat{M}_{i=3,k=1} &= \frac{1}{\sqrt2} \cos \frac{\theta}{2} |0\rangle_b {}_{a} \langle1| + \sqrt{\frac{1-D}{2}}  \sin \frac{\theta}{2} |1\rangle_b {}_{a} \langle0|, \,\,\,\,\,\quad \hat{M}_{i=3,k=2} =  \sqrt{\frac{D}{2}}  \sin \frac{\theta}{2} |0\rangle_b {}_{a} \langle0|, \\
\hat{M}_{i=4,k=1} &= - \frac{1}{\sqrt2} \cos \frac{\theta}{2} |0\rangle_b {}_{a} \langle1| + \sqrt{\frac{1-D}{2}} \sin \frac{\theta}{2} |1\rangle_b {}_{a} \langle0|, \quad \hat{M}_{i=4,k=2} = \sqrt{\frac{D}{2}}  \sin \frac{\theta}{2} |0\rangle_b {}_{a} \langle0|,
\end{aligned}
\end{equation}
which satisfies $ \sum_{i,k} \hat{M}^{\dag}_{i,k} \hat{M}_{i,k}= 1 $. 

The maximum teleportation fidelity of the conventional protocol based on conditional unitary reversal $\hat{U}^i=\{\hat{I},\hat{\sigma}_z,\hat{\sigma}_x, \hat{\sigma}_x\hat{\sigma}_z\}$ can be calculated by $F_U=\max \int d\psi \sum_i \sum_k | {}_{a}\bra{\psi} \hat{U}^i \hat{M}_{i,k} \ket{\psi}_{a}|^2$, resulting in
\begin{equation}
\label{eq:fidestand}
F_U=\frac{4-D+\cos\theta - (1-D) \cos\theta + 2\sqrt{1-D}\sin\theta}{6}.
\end{equation}
On the other hand, in MR framework, we define the reversing operators according to the recipe in Eq.~(\ref{eq:oRn}) as 
\begin{equation}
\label{eq:ordep}
\begin{aligned}
\hat{R}^{i=1} &= \hat{I} \bigg( \sqrt{1-D} \tan \frac{\theta}{2}  |0\rangle \langle0| + |1\rangle \langle1| \bigg), \\
\hat{R}^{i=2} &= \hat{\sigma}_z \bigg( \sqrt{1-D} \tan \frac{\theta}{2}  |0\rangle \langle0| + |1\rangle \langle1| \bigg), \\
\hat{R}^{i=3} &= \hat{\sigma}_x \bigg(\sqrt{1-D}  \tan \frac{\theta}{2} |0\rangle \langle0| + |1\rangle \langle1| \bigg), \\
\hat{R}^{i=4} &= \hat{\sigma}_x \hat{\sigma}_z \bigg(\sqrt{1-D} \tan \frac{\theta}{2} |0\rangle \langle0| + |1\rangle \langle1| \bigg).
\end{aligned}
\end{equation}
The teleported states obtained when the reversing operation succeeds (i.e. $j=1$) are
\begin{equation}
\label{eq:04}
\begin{aligned}
p^{i=1}_{j=1}(\psi)\rho^{i=1}_{j=1} = p^{i=2}_{j=1}(\psi)\rho^{i=2}_{j=1} &= \frac{1-D}{2} \sin^2 \frac{\theta}{2} |\psi\rangle\langle \psi| + \frac{D(1-D) |\beta|^2}{2} \sin^2 \frac{\theta}{2} \tan^2 \frac{\theta}{2}  |0\rangle \langle0|, \\
p^{i=3}_{j=1}(\psi)\rho^{i=3}_{j=1}= p^{i=4}_{j=1}(\psi)\rho^{i=4}_{j=1} &= \frac{1-D}{2} \sin^2 \frac{\theta}{2} |\psi\rangle_B {}_B \langle \psi| + \frac{D(1-D) |\alpha|^2 }{2} \sin^2 \frac{\theta}{2} \tan^2 \frac{\theta}{2} |1\rangle_B {}_B \langle1|,
\end{aligned}
\end{equation}
where the probability of successfully obtaining the teleported state $p^{i}_{j=1}(\psi) = \langle \psi| \sum_k  \hat{M}^{\dag}_{i,k} \hat{R}^{i \dag}_{j=1} \hat{R}^i_{j=1} \hat{M}_{i,k} |\psi\rangle $ is given by
\begin{equation}
\label{eq:07}
\begin{aligned}
p^{i=1}_{j=1}(\psi) = p^{i=2}_{j=1}(\psi) = \frac{1-D}{2} \sin^2 \frac{\theta}{2} + \frac{D(1-D )|\beta|^2}{2} \sin^2 \frac{\theta}{2} \tan^2 \frac{\theta}{2}, \\
p^{i=3}_{j=1}(\psi) = p^{i=4}_{j=1}(\psi) = \frac{1-D}{2} \sin^2 \frac{\theta}{2} + \frac{D(1-D )|\alpha|^2}{2} \sin^2 \frac{\theta}{2} \tan^2 \frac{\theta}{2}. 
\end{aligned}
\end{equation}

We then calculate the teleportation fidelity obtained in MR framework in Eq.~(\ref{eq:Ftele}) with the probability of obtaining the measurement outcome $i$, i.e., $ p^i(\psi) = \sum_k  \langle \psi| \hat{M}_{i,k}^{\dag} \hat{M}_{i,k} |\psi\rangle $ given by
\begin{equation}
\label{eq:rprob}
\begin{aligned}
p^{i=1}(\psi) =p^{i=2}(\psi) = \frac{|\alpha|^2 }{2} \cos^2 \frac{\theta}{2}+ \frac{ |\beta|^2}{2} \sin^2 \frac{\theta}{2}, \\
p^{i=3}(\psi) = p^{i=4}(\psi) = \frac{|\alpha|^2}{2} \sin^2 \frac{\theta}{2} + \frac{ |\beta|^2}{2} \cos^2 \frac{\theta}{2}.
\end{aligned}
\end{equation}
As a result, we obtain 
\begin{equation}
\label{eq:telefideAD2}
F_R = \int d\psi \bigg[ \frac{1 + D|\alpha|^2|\beta|^2 \tan^2 \frac{\theta}{2}}{1 + D|\beta|^2\tan^2 \frac{\theta}{2}} \bigg( |\alpha|^2 \cos^2 \frac{\theta}{2} + |\beta|^2\sin^2 \frac{\theta}{2} \bigg) + \frac{1 + D|\alpha|^2|\beta|^2\tan^2 \frac{\theta}{2}}{1 + D|\alpha|^2\tan^2 \frac{\theta}{2}} \bigg( |\alpha|^2\sin^2 \frac{\theta}{2} + |\beta|^2\cos^2 \frac{\theta}{2}\bigg)
\bigg].
\end{equation}
We compare $F_R$ with $F_U$ in Eq.~(\ref{eq:fidestand}) by changing  $0\leq \theta \leq \pi/2$ and the degree of decoherence $0 \leq D \leq 1$ in Fig.~\ref{figs1}(a), which clearly shows that $F_R \geq F_U$ in all regions.

\begin{figure*}[b]
\centering
\includegraphics[width=5.0 in]{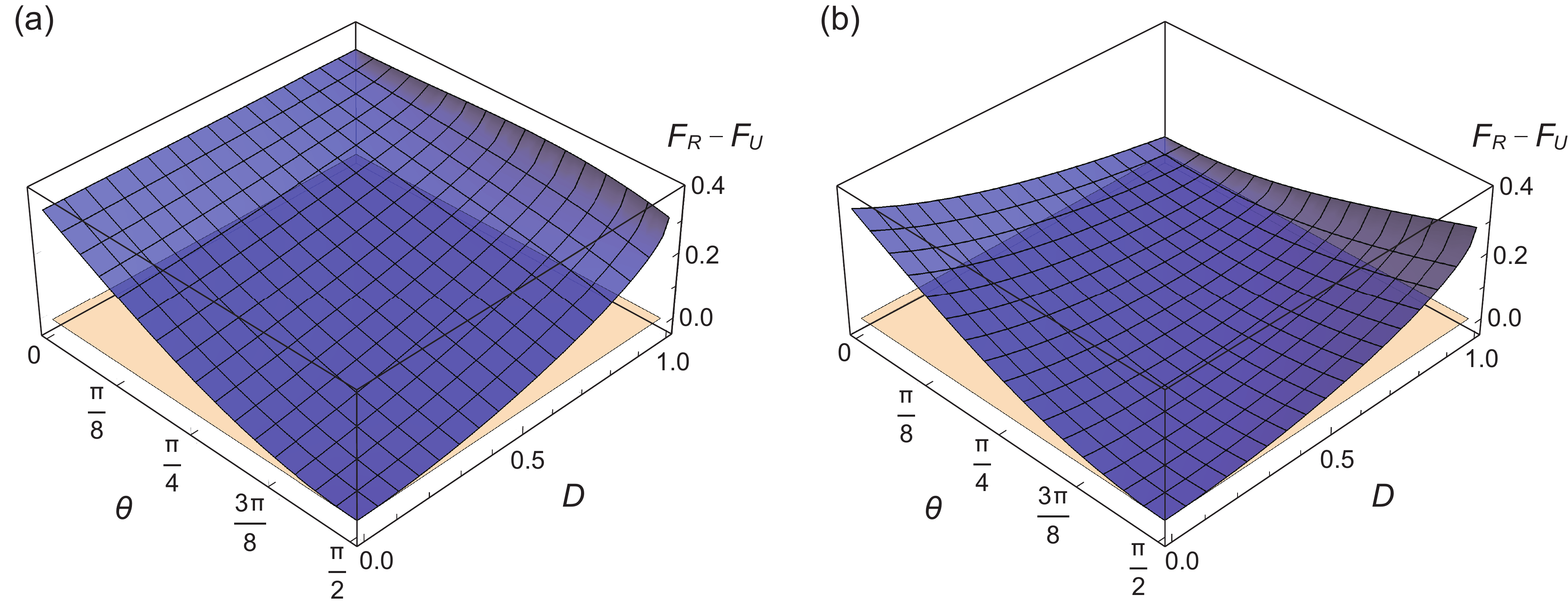}
\caption{\textbf{Teleportation fidelity under the amplitude damping decoherence.} Teleportation fidelities are plotted by changing $D$ and $\theta$ under the amplitude damping in modes (a) $b$ and (b) $\bar{a}$. In all regions, the optimal reversing operation yields a higher teleportation fidelity compared to the conventional protocol based on conditional unitary operation, i.e., $F_R-F_U \geq 0$ always holds. The yellow plane represents $F_R-F_U=0$. 
These results show that the teleportation protocol optimized within our general framework overcomes the limit of conventional quantum teleportation protocols.
}
\label{figs1}
\end{figure*}

\item[(ii)] {\em Decoherence in mode $\bar{a}$} -- Let us now consider the case when decoherence occurs in mode $\bar{a}$, assuming for simplicity that decoherence does not occur on mode $b$. The mode $\bar{a}$ of the channel $\ket{\Psi}_{\bar{a}b}$ then experience the amplitude damping noise with operators $ \hat{E}_{1} \equiv |0\rangle \langle 0| +  \sqrt{1-\text{D}} |1\rangle \langle 1|$ and $ \hat{E}_{2} \equiv \sqrt{\text{D}} |0\rangle \langle 1| $. In this case, the maximum teleportation fidelity obtained in conventional protocols is the same as Eq.~(\ref{eq:fidestand}). On the other hand, within our general framework, we can find an appropriate joint measurement basis for a given decoherence model. For example, we can choose a joint measurement basis as
\begin{equation}
\begin{aligned}
\ket{W_1}&=\frac{1}{\sqrt{2-D}}(\sqrt{1-D}\ket{00}+\ket{11}),\\
\ket{W_2}&=\frac{1}{\sqrt{2-D}}(\ket{00}-\sqrt{1-D}\ket{11}),\\
\ket{W_3}&=\frac{1}{\sqrt{2-D}}(\ket{01}+\sqrt{1-D}\ket{10}),\\
\ket{W_4}&=\frac{1}{\sqrt{2-D}}(\sqrt{1-D}\ket{01}-\ket{10}),
\end{aligned}
\end{equation}
so that the corresponding effective quantum measurement is given by $ \hat{M}_{i,k} = {}_{a\bar{a}} \langle W_i |  \hat{E}_{k,\bar{a}} | \Psi\rangle_{\bar{a}b} $ as 
\begin{equation}
\begin{aligned}
\hat{M}_{i=1,k=1} &= \sqrt{\frac{1-D}{2-D}} \bigg( \cos \frac{\theta}{2} |0\rangle_b {}_a \langle0| + \sin \frac{\theta}{2} |1\rangle_b {}_a \langle1| \bigg), ~~~ \hat{M}_{i=1,k=2} = \sqrt{\frac{D(1-D)}{2-D}} \sin \frac{\theta}{2} |1\rangle_b {}_a \langle0|, \\
\hat{M}_{i=2,k=1}&= \frac{1}{\sqrt{2-D}} \bigg( \cos \frac{\theta}{2} |0\rangle_b {}_a \langle0| - (1-D)\sin \frac{\theta}{2} |1\rangle_b {}_a \langle1| \bigg), ~~~ \hat{M}_{i=2,k=2} =  \sqrt{\frac{D}{2-D}} \sin \frac{\theta}{2} |1\rangle_b {}_a \langle0|, \\
\hat{M}_{i=3,k=1} &=  \sqrt{\frac{1-D}{2-D}} \bigg( \cos \frac{\theta}{2} |0\rangle_b {}_a \langle1| + \sin \frac{\theta}{2} |1\rangle_b {}_a \langle0| \bigg), ~~~ \hat{M}_{i=3,k=2} = \sqrt{\frac{D(1-D)}{2-D}} \sin \frac{\theta}{2} |1\rangle_b {}_a \langle1|, \\
\hat{M}_{i=4,k=1} &= \frac{-1}{\sqrt{2-D}} \bigg( \cos \frac{\theta}{2} |0\rangle_b {}_a \langle1| - (1-D)\sin \frac{\theta}{2} |1\rangle_b {}_a \langle0| \bigg), ~~~ \hat{M}_{i=4,k=2} = \sqrt{\frac{D}{2-D}} \sin \frac{\theta}{2} |1\rangle_b {}_a \langle1|. 
\end{aligned}
\end{equation}
The probability of obtaining each outcome $i$ is then calculated by $ p^i(\psi) = \sum_k  \langle \psi| \hat{M}^{\dag}_{i,k} \hat{M}_{i,k} |\psi\rangle $ as
\begin{equation}
\label{eq:probout2}
\begin{aligned}
 p^{i=1}(\psi) &=\frac{1-D}{2-D} \bigg( |\alpha|^2\cos^2  \frac{\theta}{2}  + |\beta|^2\sin^2 \frac{\theta}{2}  \bigg) + \frac{D(1-D)}{2-D}  |\alpha|^2 \sin^2 \frac{\theta}{2}, \\
 p^{i=2}(\psi) &= \frac{1}{2-D} \bigg( |\alpha|^2\cos^2  \frac{\theta}{2} + (1-D)^2 |\beta|^2 \sin^2 \frac{\theta}{2} \bigg) + \frac{D}{2-D}  |\alpha|^2\sin^2 \frac{\theta}{2}, \\
 p^{i=3}(\psi) &=\frac{1-D}{2-D} \bigg( |\beta|^2\cos^2  \frac{\theta}{2} +  |\alpha|^2  \sin^2 \frac{\theta}{2}\bigg) + \frac{D(1-D)}{2-D}|\beta|^2 \sin^2 \frac{\theta}{2} , \\
 p^{i=4}(\psi) &= \frac{1}{2-D} \bigg( |\beta|^2\cos^2  \frac{\theta}{2} + (1-D)^2 |\alpha|^2 \sin^2 \frac{\theta}{2}  \bigg) + \frac{D}{2-D}  |\beta|^2\sin^2 \frac{\theta}{2}. 
\end{aligned}
\end{equation}

The results for the outcome $i=1$ and $i=3$ are then selected, and after applying an appropriate reversing operation with $\hat{R}^i=\hat{U}^i(\tan (\theta/2) \ket{0}\bra{0} + \ket{1}\bra{1})$ where $\hat{U}^i=\{\hat{I},\hat{\sigma}_z,\hat{\sigma}_x,\hat{\sigma}_x\hat{\sigma}_z\}$, the output states in mode $b$ are obtained as
\begin{equation}
\begin{aligned}
p^{i=1}_b(\psi)\rho^{i=1}_b &=  \frac{1-D}{2-D} \sin^2 \frac{\theta}{2} |\psi\rangle \langle \psi| + \frac{D(1-D)}{2-D}  |\alpha|^2 \sin^2 \frac{\theta}{2} |1\rangle \langle1|, \\
p^{i=3}_b(\psi)\rho^{i=3}_b&=  \frac{1-D}{2-D} \sin^2 \frac{\theta}{2} |\psi\rangle \langle \psi| + \frac{D(1-D)}{2-D} |\beta|^2 \sin^2 \frac{\theta}{2}  |0\rangle \langle0|,
\end{aligned}
\end{equation}
where 
\begin{equation}
\label{eq:probra}
\begin{aligned}
p^{i=1}_b(\psi) &= \frac{(1-D)(1+D|\alpha|^2)}{2-D} \sin^2 \frac{\theta}{2} ,\\
p^{i=3}_b(\psi) &= \frac{(1-D)(1+D|\beta|^2)}{2-D} \sin^2 \frac{\theta}{2} . 
\end{aligned}
\end{equation}
We can then calculate the teleportation fidelity as 
\begin{equation}
F_R=\int d\psi \bra{\psi} \frac{p^{i=1}(\psi)\rho^{i=1}_b+p^{i=3}(\psi)\rho^{i=3}_b}{p^{i=1}(\psi)+p^{i=3}(\psi)} \ket{\psi}.
\end{equation}
As shown in Fig.~\ref{figs1}(b), the teleportation protocol optimized within our framework outperforms the conventional protocols.

\item[(iii)] {\em Other forms of decoherence} -- We can also similarly find the optimal protocols to maximize the teleportation fidelity in the presence of other types of decoherence, dephasing and depolarizing noises as described by \cite{QIQM}
\begin{equation}
\label{eq:othernoise}
\begin{aligned}
{\rm (Dephasing):}&~\hat{E}_{1} = |0\rangle \langle 0| +  \sqrt{1-\text{D}} |1\rangle \langle 1|,~\hat{E}_{2} = \sqrt{\text{D}} |1\rangle \langle 1| \\
{\rm (Depolarizing):}&~ \hat{E}_{1} = \sqrt{1-3D/4}\hat{I},~\hat{E}_{2} = \sqrt{D/4}\hat{\sigma}_x,~\hat{E}_{3} = \sqrt{D/4}\hat{\sigma}_y,~\hat{E}_{4} = \sqrt{D/4}\hat{\sigma}_z.
\end{aligned}
\end{equation}
We then define the corresponding quantum measurement $ \hat{M}_{i,k} = {}_{a\bar{a}} \langle W_i |  \hat{E}_{k,b} | \Psi\rangle_{\bar{a}b} $ (on mode $b$) as 
\begin{equation}
\begin{aligned}
{\rm (Dephasing):}&~\\
\hat{M}_{i=1,k=1} &= \frac{1}{\sqrt2} \cos \frac{\theta}{2} |0\rangle_b {}_{a} \langle0| + \sqrt{\frac{1-D}{2}} \sin \frac{\theta}{2} |1\rangle_b {}_{a} \langle1|, \,\,\,\,\,\quad \hat{M}_{i=1,k=2} = \sqrt{\frac{D}{2}} \sin \frac{\theta}{2} |1\rangle_b {}_{a} \langle1| \\
\hat{M}_{i=2,k=1} &= \frac{1}{\sqrt2} \cos \frac{\theta}{2} |0\rangle_b {}_{a} \langle0| - \sqrt{\frac{1-D}{2}}  \sin \frac{\theta}{2} |1\rangle_b {}_{a} \langle1|, \,\,\,\,\,\quad \hat{M}_{i=2,k=2} = - \sqrt{\frac{D}{2}}  \sin \frac{\theta}{2} |1\rangle_b {}_{a} \langle1| \\
\hat{M}_{i=3,k=1} &= \frac{1}{\sqrt2} \cos \frac{\theta}{2} |0\rangle_b {}_{a} \langle1| + \sqrt{\frac{1-D}{2}}  \sin \frac{\theta}{2} |1\rangle_b {}_{a} \langle0|, \,\,\,\,\,\quad \hat{M}_{i=3,k=2} =  \sqrt{\frac{D}{2}}  \sin \frac{\theta}{2} |1\rangle_b {}_{a} \langle0| \\
\hat{M}_{i=4,k=1} &= - \frac{1}{\sqrt2} \cos \frac{\theta}{2} |0\rangle_b {}_{a} \langle1| + \sqrt{\frac{1-D}{2}} \sin \frac{\theta}{2} |1\rangle_b {}_{a} \langle0|, \quad \hat{M}_{i=4,k=2} = \sqrt{\frac{D}{2}}  \sin \frac{\theta}{2} |1\rangle_b {}_{a} \langle0|
\end{aligned}
\end{equation}
\begin{equation}
\begin{aligned}
{\rm (Depolarizing):}&~\\
\hat{M}_{i,k=1} &= \frac{1}{\sqrt2}\sqrt{1-\frac{3D}{4}} \Big( \cos \frac{\theta}{2} |0\rangle_b {}_{a} \langle0| +  \sin \frac{\theta}{2} |1\rangle_b {}_{a} \langle1| \Big)\hat{U}^i \\
\hat{M}_{i,k=2} &= \frac{1}{\sqrt2}\sqrt{\frac{3D}{4}} \Big( \cos \frac{\theta}{2} |0\rangle_b {}_{a} \langle1| +  \sin \frac{\theta}{2} |1\rangle_b {}_{a} \langle0| \Big)\hat{U}^i \\
\hat{M}_{i,k=3} &= \frac{i}{\sqrt2}\sqrt{\frac{3D}{4}} \Big( \cos \frac{\theta}{2} |0\rangle_b {}_{a} \langle1| -  \sin \frac{\theta}{2} |1\rangle_b {}_{a} \langle0| \Big)\hat{U}^i  \\
\hat{M}_{i,k=4} &= \frac{1}{\sqrt2}\sqrt{\frac{3D}{4}} \Big( \cos \frac{\theta}{2} |0\rangle_b {}_{a} \langle0| -  \sin \frac{\theta}{2} |1\rangle_b {}_{a} \langle1| \Big)\hat{U}^i
\end{aligned}
\end{equation}
where $\hat{U}^i=\{\hat{I},\hat{\sigma}_z,\hat{\sigma}_x, \hat{\sigma}_x\hat{\sigma}_z\}$, which satisfies $ \sum_{i,k} \hat{M}^{\dag}_{i,k} \hat{M}_{i,k}= 1 $. 

The maximum teleportation fidelity of the conventional protocol based on conditional unitary reversal $\hat{U}^i=\{\hat{I},\hat{\sigma}_z,\hat{\sigma}_x, \hat{\sigma}_x\hat{\sigma}_z\}$ can be obtained as
\begin{equation}
\label{eq:othernoiseU}
\begin{aligned}
{\rm (Dephasing):}&~F_U=\frac{2 + \sqrt{1-D}\sin\theta}{3} \\
{\rm (Depolarizing):}&~ F_U=\frac{4-D + 2\sqrt{1-D}\sin\theta}{6}
\end{aligned}
\end{equation}

\begin{figure*}[ht]
\centering
\includegraphics[width=5.0 in]{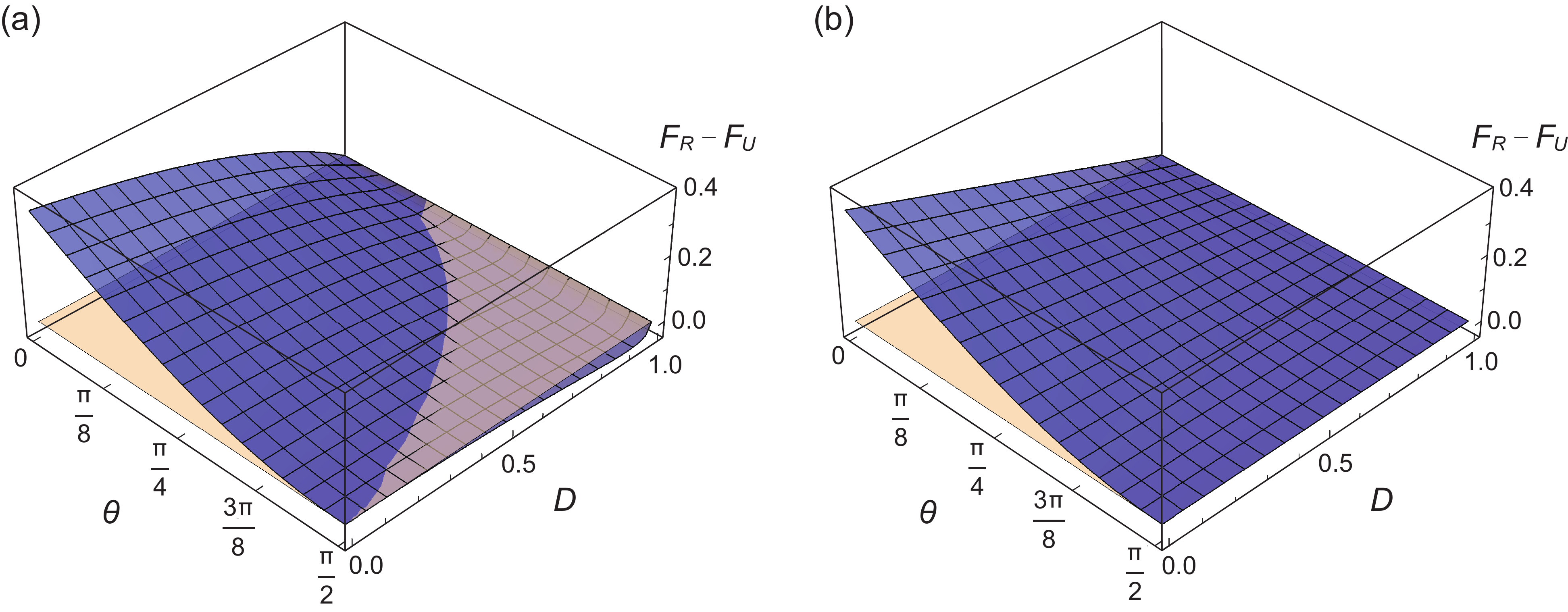}
\caption{\textbf{Teleportation fidelity under the dephasing and the depolarizing decoherence.} Teleportation  fidelities are plotted by changing $D$ and $\theta$ for (a) the dephasing and (b) the depolarization in mode $b$. For the dephasing, (a) the optimal reversing operation yields a higher teleportation fidelity $F_R$ compared to the conventional protocols $F_U$ based on conditional unitary operation in most cases. 
 For the depolarization, (b) $F_R$ is always larger than $F_U$, i.e.,  $F_R-F_U \geq 0$ holds. The yellow plane represents $F_R-F_U=0$. 
 These results show that our general framework enables us to teleport quantum states under decoherence beyond the reach of the conventional protocols.
}
\label{figs2}
\end{figure*}

In MR framework, we can define the reversing operators. The reversing operators for the dephasing noise are given in the same form with (\ref{eq:ordep}), while those for the depolarizing noise are defined as 
\begin{equation}
\hat{R}^i=\hat{U}^i \bigg( \tan \frac{\theta}{2} \ket{0}\bra{0}+\ket{1}\bra{1} \bigg).
\end{equation}

We then calculate the teleportation fidelity $F_R$ obtained by the reversing operation likewise we did for the damping noise, resulting in
\begin{align}
\label{eq:telefideDPs}
&{\rm (Dephasing):} F_R = \int d\psi \bigg[ \frac{1-D+D|\beta|^4}{1-D+D|\beta|^2} \bigg( |\alpha|^2 \cos^2 \frac{\theta}{2} + |\beta|^2\sin^2 \frac{\theta}{2} \bigg) + \frac{1-D+D|\alpha|^4}{1-D+D|\alpha|^2} \bigg( |\alpha|^2\sin^2 \frac{\theta}{2} + |\beta|^2\cos^2 \frac{\theta}{2}\bigg)
\bigg],\\
&{\rm (Depolarizing):}~~F_R=1-\frac{D}{2}.
\end{align}

We compare $F_R$ with $F_U$ in Eq.~(\ref{eq:othernoiseU}) by changing  $0\leq \theta \leq \pi/2$ and the degree of decoherence $0 \leq D \leq 1$ resulting in Fig.~\ref{figs2}(a) and (b). As a result, we can find that MR framework allows us to optimize quantum teleportation protocol such that $F_{\rm tele}=\max[F_U,F_R]$ even via a severely decohered arbitrary quantum channel beyond the reach of the conventional protocols as we plot in Fig.~\ref{fig:fig2}.

\end{itemize}

\section{Application to teleportation-based functionalities for quantum network}
\label{asec:app}

In this section, we apply MR framework to teleportation-based functionalities that are potentially useful in building scalable quantum architectures or quantum networks as illustrated in Fig~\ref{fig:App}.

\begin{itemize}

\item[(a)] Multipartite communication: Suppose that Alice teleports a qubit $\ket{\psi}_{a}$ to Charlie by cooperating with Bob. Alice, Bob, and Charlie are connected in quantum network via a GHZ-type entangled state $\ket{\Psi}_{\bar{a}bc}=\cos(\theta/2)\ket{000}+\sin(\theta/2)\ket{111}$ ($0 \leq \theta \leq \pi/2$) as illustrated in Fig.~\ref{fig:App}(a). For this, Alice performs a joint measurement in Bell basis $\ket{v_{i}}$ (with outcomes $i=1,2,3,4$) between the input qubit mode $\bar{a}$ and the one mode of the entangled channel $a$. Bob performs a projection measurement (with outcomes $i'=1,2$) on mode $b$ of the channel in the basis $\ket{w_1}=\cos(\phi/2)\ket{0}+\sin(\phi/2)\ket{1}$, $\ket{w_2}=-\sin(\phi/2)\ket{0}+\cos(\phi/2)\ket{1}$ where $0 \leq \phi \leq \pi/2$. Alice and Bob send their measurement outcomes $i$ and $i'$, respectively, to Charlie via classical channel. 
The measurement operator of the teleportation in MR framework can be written by
\begin{equation}
\hat{M}_{i,i'}\equiv {}_b\bra{w_{i'}}{}_{a\bar{a}}\bra{v_{i}}\cdot \ket{\Psi}_{\bar{a}bc}.
\end{equation}
If we assume $\theta \geq \phi$, the measurement operator for all possible measurement outcomes $i=1,2,3,4$ and $i'=1,2$ are given by
\begin{equation}
\begin{aligned}
\hat{M}_{i=1,i'=1}&=\frac{1}{\sqrt{2}}\Big(\cos{\frac{\theta}{2}}\cos{\frac{\phi}{2}} \ket{0}\bra{0}+\sin{\frac{\theta}{2}}\sin{\frac{\phi}{2}}\ket{1}\bra{1}\Big)\\
\hat{M}_{i=1,i'=2}&=\frac{1}{\sqrt{2}} \sigma_x \sigma_z \Big(\sin{\frac{\theta}{2}}\cos{\frac{\phi}{2}} \ket{0}\bra{0}+\cos{\frac{\theta}{2}}\sin{\frac{\phi}{2}}\ket{1}\bra{1}\Big) \sigma_x \\
\hat{M}_{i=2,i'=1}&=\frac{1}{\sqrt{2}} \sigma_z \Big(\cos{\frac{\theta}{2}}\cos{\frac{\phi}{2}} \ket{0}\bra{0}+\sin{\frac{\theta}{2}}\sin{\frac{\phi}{2}}\ket{1}\bra{1}\Big)\\
\hat{M}_{i=2,i'=2}&=\frac{1}{\sqrt{2}} \sigma_x \Big(\sin{\frac{\theta}{2}}\cos{\frac{\phi}{2}} \ket{0}\bra{0}+\cos{\frac{\theta}{2}}\sin{\frac{\phi}{2}}\ket{1}\bra{1}\Big) \sigma_x \\
\hat{M}_{i=3,i'=1}&=\frac{1}{\sqrt{2}}\Big(\cos{\frac{\theta}{2}}\cos{\frac{\phi}{2}} \ket{0}\bra{0}+\sin{\frac{\theta}{2}}\sin{\frac{\phi}{2}}\ket{1}\bra{1}\Big) \sigma_x\\
\hat{M}_{i=3,i'=2}&=\frac{1}{\sqrt{2}} \sigma_x \sigma_z \Big(\sin{\frac{\theta}{2}}\cos{\frac{\phi}{2}} \ket{0}\bra{0}+\cos{\frac{\theta}{2}}\sin{\frac{\phi}{2}}\ket{1}\bra{1}\Big)\\
\hat{M}_{i=4,i'=1}&=\frac{1}{\sqrt{2}}\Big(\cos{\frac{\theta}{2}}\cos{\frac{\phi}{2}} \ket{0}\bra{0}+\sin{\frac{\theta}{2}}\sin{\frac{\phi}{2}}\ket{1}\bra{1}\Big) \sigma_x \sigma_z \\
\hat{M}_{i=4,i'=2}&=\frac{1}{\sqrt{2}} \sigma_x \Big(\sin{\frac{\theta}{2}}\cos{\frac{\phi}{2}} \ket{0}\bra{0}+\cos{\frac{\theta}{2}}\sin{\frac{\phi}{2}}\ket{1}\bra{1}\Big).
\end{aligned}
\end{equation}
We can calculate the amount of information obtained by Alice through this measurement as
\begin{equation}
\label{eq:gainA1}
{\cal G}^{\rm max}_{\rm Alice}=\frac{1}{3}\bigg(1+\sin^2{\frac{\phi}{2}}\bigg)
\end{equation}
by following the definition of the information gain of quantum measurement \cite{Bana}. Note that a teleportation protocol conventionally based on unitary reversal yields the teleportation fidelity ${\cal F}_U=(2+\sin{\theta}\sin{\phi})/3$. 
By contrast, in MR framework, we can find the optimal reversing operators for each measurement outcomes as
\begin{equation}
\begin{aligned}
\hat{R}^{i=1,i'=1}&=\frac{1}{\sqrt{2}}\Big(\tan{\frac{\theta}{2}}\tan{\frac{\phi}{2}} \ket{0}\bra{0}+\ket{1}\bra{1}\Big)\\
\hat{R}^{i=1,i'=2}&=\frac{1}{\sqrt{2}} \sigma_x \sigma_z \Big(\cot{\frac{\theta}{2}}\tan{\frac{\phi}{2}} \ket{0}\bra{0}+\ket{1}\bra{1}\Big) \sigma_x \\
\hat{R}^{i=2,i'=1}&=\frac{1}{\sqrt{2}} \sigma_z \Big(\tan{\frac{\theta}{2}}\tan{\frac{\phi}{2}} \ket{0}\bra{0}+\ket{1}\bra{1}\Big)\\
\hat{R}^{i=2,i'=2}&=\frac{1}{\sqrt{2}} \sigma_x  \Big(\cot{\frac{\theta}{2}}\tan{\frac{\phi}{2}} \ket{0}\bra{0}+\ket{1}\bra{1}\Big) \sigma_x \\
\hat{R}^{i=3,i'=1}&=\frac{1}{\sqrt{2}} \sigma_x \Big(\tan{\frac{\theta}{2}}\tan{\frac{\phi}{2}} \ket{0}\bra{0}+\ket{1}\bra{1}\Big)\\
\hat{R}^{i=3,i'=2}&=\frac{1}{\sqrt{2}} \sigma_z \Big(\cot{\frac{\theta}{2}}\tan{\frac{\phi}{2}} \ket{0}\bra{0}+\ket{1}\bra{1}\Big) \sigma_x\\
\hat{R}^{i=4,i'=1}&=\frac{1}{\sqrt{2}} \sigma_z \sigma_x \Big(\tan{\frac{\theta}{2}}\tan{\frac{\phi}{2}} \ket{0}\bra{0}+\ket{1}\bra{1}\Big) \\
\hat{R}^{i=4,i'=2}&=\frac{1}{\sqrt{2}} \Big(\cot{\frac{\theta}{2}}\tan{\frac{\phi}{2}} \ket{0}\bra{0}+\ket{1}\bra{1}\Big) \sigma_x,
\end{aligned}
\end{equation}
to accomplish $\hat{R}^{i,i'}\hat{M}_{i,i'}\ket{\psi}\propto\ket{\psi}$. Therefore, by applying these optimal reversing operation, it is possible to faithfully recover the input qubit $\ket{\psi}$ at Charlie's location, i.e., ${\cal F}_{\rm tele}=1$. The maximum success probability of the teleportation protocol can be evaluated as 
\begin{equation}
{\cal P}^{\rm max}_{\rm tele}=2\sin^2{\frac{\phi}{2}}.
\end{equation}
It is straightforward to see that ${\cal P}^{\rm max}_{\rm tele}$ and ${\cal G}^{\rm max}_{\rm Alice}$ in Eq.~(\ref{eq:gainA1}) saturate the upper bound in Eq.~(\ref{eq:balance}). 
 
\item[(b)] Entanglement transmission: Suppose that Alice teleports an arbitrary two qubit state $\ket{\psi}_{ac}=\alpha\ket{00}+\beta\ket{10}+\gamma\ket{01}+\delta\ket{11}$ to Bob via two entangled pairs: one is in an arbitrary entangled state $\ket{\Psi}_{\bar{a}b}=\cos(\theta/2)\ket{00}+\sin(\theta/2)\ket{11}$ ($0 \leq \theta \leq \pi/2$) and the other is a maximally entangled state $\ket{\Psi}_{\bar{c}d}=(\ket{00}+\ket{11})/\sqrt{2}$ as illustrated in Fig.~\ref{fig:App}(b). We also assume that the first joint measurement between mode $a$ and $\bar{a}$ is performed in the Bell basis $\ket{v_{i}}$ ($i=1,2,3,4$), while the other joint measurement between mode $c$ and $\bar{c}$ ($i'=1,2,3,4$) is performed in the basis
\begin{equation}
\begin{aligned}
\ket{v'_1}&=\cos{\frac{\phi}{2}}\ket{00}+\sin{\frac{\phi}{2}}\ket{11}\\
\ket{v'_2}&=\sin{\frac{\phi}{2}}\ket{00}-\cos{\frac{\phi}{2}}\ket{11}\\
\ket{v'_3}&=\cos{\frac{\phi}{2}}\ket{01}+\sin{\frac{\phi}{2}}\ket{10}\\
\ket{v'_4}&=\sin{\frac{\phi}{2}}\ket{01}-\cos{\frac{\phi}{2}}\ket{10}.
\end{aligned}
\end{equation}
In MR framework, the quantum measurement for all measurement outcomes $i=1,2,3,4$ and $i'=1,2,3,4$ is defined by
\begin{equation}
\hat{M}_{i,i'}\equiv {}_{c\bar{c}}\bra{v_{i'}}{}_{a\bar{a}}\bra{v_{i}}\cdot \ket{\Psi}_{\bar{a}b}\ket{\Psi}_{\bar{c}d}.
\end{equation}
For example, when the outcome is $i=1$ and $i'=2$, the corresponding measurement operator is given by
\begin{equation}
\hat{M}_{i=1,i'=2}=\frac{1}{2}\Big(\cos{\frac{\theta}{2}}\ket{0}_{ba}\bra{0}+\sin{\frac{\theta}{2}}\ket{1}_{ba}\bra{1}\Big)\otimes \Big(\sin{\frac{\phi}{2}}\ket{0}_{dc}\bra{0}-\cos{\frac{\phi}{2}}\ket{1}_{dc}\bra{1}\Big),
\end{equation}
and its optimal reversing operator is
\begin{equation}
\hat{R}^{i=1,i'=2}=\Big(\tan{\frac{\theta}{2}}\ket{0}_{b}\bra{0}+\ket{1}_{b}\bra{1}\Big)\otimes \Big(-\tan{\frac{\phi}{2}}\ket{0}_{d}\bra{0}-\ket{1}_{d}\bra{1}\Big).
\end{equation}
While a conventional teleportation based on unitary reversal can achieve at best ${\cal F}_U=(1+(1+\sin\theta)(1+\sin\phi))$, the optimized protocol in MR framework enables a faithful teleportation, i.e., ${\cal F}_{\rm tele}=1$ with the probability
\begin{equation}
{\cal P}^{\rm max}_{\rm tele}=4\sin^2{\frac{\theta}{2}}\sin^2{\frac{\phi}{2}},
\end{equation}
which satisfies the performance limit in Eq.~(\ref{eq:balance}) together with the amount of information gain by Alice that can be evaluated as ${\cal G}^{\rm max}_{\rm Alice}=(1/5)(1+\cos^2(\theta/2)\cos^2(\phi/2))$. Note that the upper bound of the trade-off relation is not saturated in this scenario since the input state $\ket{\psi}_{ac}$ is in $d=4$ dimensional Hilbert space in which only a specific set of quantum measurements can saturate the bound. The saturation condition for arbitrary $d$-dimensional quantum measurements is given in Ref.~\cite{Cheong12}.

\item[(c)] One-way quantum repeater: Alice transmits a qubit over long distance to David with the help of teleportation-based quantum repeaters at intermediated nodes (Bob and Charlie) as illustrated in Fig.~\ref{fig:App}(c). Let us assume that Bob performs a joint measurement in the Bell basis $\ket{v_i}$ ($i=1,2,3,4$) between the qubit Alice has sent $\ket{\psi}_a$ and one of the entangled qubit pair prepared in Bob's node $\ket{\Psi}_{bc}=\cos(\theta/2)\ket{00}+\sin(\theta/2)\ket{11}$, and transmits the remaining qubit to Charlie. Then, Charlie performs a joint measurement in the basis ($i'=1,2,3,4$)
\begin{equation}
\begin{aligned}
\ket{v'_1}&=\cos{\frac{\phi}{2}}\ket{00}+\sin{\frac{\phi}{2}}\ket{11}\\
\ket{v'_2}&=\sin{\frac{\phi}{2}}\ket{00}-\cos{\frac{\phi}{2}}\ket{11}\\
\ket{v'_3}&=\cos{\frac{\phi}{2}}\ket{01}+\sin{\frac{\phi}{2}}\ket{10}\\
\ket{v'_4}&=\sin{\frac{\phi}{2}}\ket{01}-\cos{\frac{\phi}{2}}\ket{10}
\end{aligned}
\end{equation}
between the qubit from Bob and one of the maximally entangled qubits, and transmits the remaining qubit to David. In MR framework, the quantum measurement representation of the protocol (for all measurement outcomes $i=1,2,3,4$ and $i'=1,2,3,4$) is defined as 
\begin{equation}
\hat{M}_{i,i'}\equiv {}_{bc}\bra{v_{i'}}{}_{a\bar{a}}\bra{v_{i}}\cdot \ket{\Psi}_{\bar{a}b}\ket{\Psi}_{cd}.
\end{equation}
For example, when the outcome of Bob's joint measurement is $i=2$ and the outcome of Charlie's joint measurement is $i'=1$, the corresponding measurement operator is written by
\begin{equation}
\hat{M}_{i=2,i'=1}=\frac{1}{2}\Big(\cos{\frac{\theta}{2}}\cos{\frac{\phi}{2}}\ket{0}_{da}\bra{0}-\sin{\frac{\theta}{2}}\sin{\frac{\phi}{2}}\ket{1}_{da}\bra{1}\Big),
\end{equation}
and its optimal reversing operator is
\begin{equation}
\hat{R}^{i=2,i'=1}=\tan{\frac{\theta}{2}}\tan{\frac{\phi}{2}}\ket{0}_{b}\bra{0}-\ket{1}_{b}\bra{1}.
\end{equation}
In this scenario, while the maximum teleportation fidelity obtained by conventional protocols based on unitary reversal is ${\cal F}_U=(2+\sin{\theta}\sin{\phi})/3$, the optimize protocol in MR framework enables a faithful transmission, i.e., ${\cal F}_{\rm tele}=1$ over long distance with the success probability 
\begin{equation}
{\cal P}^{\rm max}_{\rm tele}=2\sin^2\Big({\frac{\min[\theta,\phi]}{2}}\Big).
\end{equation}
This saturates the upper bound in Eq.~(\ref{eq:balance}) together with the amount of information gain by Alice evaluated as ${\cal G}^{\rm max}_{\rm Alice}=(1/3)(1+\cos^2(\min[\theta,\phi]/2))$.

\end{itemize}

We note that, likewise aforementioned scenarios, any multipartite quantum teleportation including multiple senders, intermediators, and receivers in quantum network can be optimized in MR framework so that an unknown quantum state in quantum network can be transmitted with the unit fidelity ${\cal F}_{\rm tele}=1$ with the maximum success probability satisfying (or saturating) the performance limit. It would be generally useful for any quantum communication protocols in complicated quantum network.

\end{widetext}

\end{document}